\documentclass[journal]{IEEEtran}

\usepackage{subfigure}
\usepackage{setspace}
\usepackage{multirow} 
\usepackage{algorithm}
\usepackage{algorithmic}
\usepackage{amsmath}
\usepackage{amssymb}
\usepackage{latexsym}
\usepackage{multirow}
\usepackage{epsfig}
\usepackage{graphics}
\usepackage{graphicx}
\usepackage{mathrsfs}
\usepackage{subfigure}
\usepackage{mathrsfs}
\usepackage{bbding}
\usepackage{cite}
\usepackage{color}
\usepackage{xcolor}
\usepackage{booktabs}
\usepackage{amssymb,mathrsfs,amsmath}
\newcommand{\bm}[1]{\mbox{\boldmath{$#1$}}}
\usepackage{amsthm}
\theoremstyle{plain}
\newtheorem{prop}{Proposition}
\newtheorem{rem}{Remark}
\usepackage{flushend}

\usepackage{hyperref}

\allowdisplaybreaks [4]

\begin{document}	

\title{Integrating Movable Antennas and Intelligent Reflecting Surfaces for Coverage Enhancement}
\author{Ying~Gao, Qingqing~Wu, Weidong~Mei, Guangji~Chen, Wen~Chen, and Ziyuan~Zheng\vspace{-5mm} 
\thanks{Y.~Gao, Q.~Wu, W.~Chen, and Z.~Zheng are with the Department of Electronic Engineering, Shanghai Jiao Tong University, Shanghai 201210, China (e-mail: yinggao@sjtu.edu.cn; qingqingwu@sjtu.edu.cn; wenchen@sjtu.edu.cn; zhengziyuan2024@sjtu.edu.cn). W.~Mei is with the National Key Laboratory of Wireless Communications, University of Electronic Science and Technology of China, Chengdu 611731, China (e-mail: wmei@uestc.edu.cn). G.~Chen is with the School of Electronic and Optical Engineering, Nanjing University of Science and Technology, Nanjing 210094, China and also with National Mobile Communications Research Laboratory, Southeast University (email: guangjichen@njust.edu.cn).}}

\maketitle

\begin{abstract}
	This paper investigates an intelligent reflecting surface (IRS)-aided movable antenna (MA) system, where multiple IRSs cooperate with a multi-MA base station to extend wireless coverage to multiple target areas. The objective is to maximize the worst-case signal-to-noise ratio (SNR) across all locations within these areas through joint optimization of MA positions, IRS phase shifts, and transmit beamforming. 
	To achieve this while balancing the performance-cost trade-off, we propose three coverage-enhancement schemes: the \emph{area-adaptive MA-IRS} scheme, where both the MA positions and IRS phase shifts are adaptively adjusted for each target area; the \emph{area-adaptive MA-staIRS} scheme, where only the MA positions are adjusted, while the IRS phase shifts remain unchanged after initial configuration (with \emph{staIRS} denoting static IRSs); and the \emph{shared MA-staIRS} scheme, where a common MA placement and static IRS configuration are applied across all areas. 
	These schemes lead to challenging non-convex optimization problems with implicit objective functions, which are difficult to solve optimally. To address these problems, we propose a general algorithmic framework that can be applied to solve each problem efficiently albeit suboptimally. Simulation results demonstrate that: 1) the proposed MA-based schemes consistently outperform their fixed-position antenna (FPA)-based counterparts under both area-adaptive and static IRS configurations, with the area-adaptive MA-IRS scheme achieving the highest worst-case SNR; 2) as transmit antennas are typically far fewer than IRS elements, the area-adaptive MA-staIRS scheme may underperform the baseline FPA scheme with area-adaptive IRSs in terms of the worst-case SNR, but a modest increase in antenna number can reverse this trend; 3) under a fixed total cost, the optimal MA-to-IRS-element ratio for the worst-case SNR maximization is empirically found to be proportional to the reciprocal of their unit cost ratio.  
\end{abstract}


\vspace{-2mm}
\section{Introduction}
In recent decades, wireless communication has undergone revolutionary advancements, with multiple-input multiple-output (MIMO) technology emerging as a key innovation for enhancing spectral efficiency and reliability via spatial diversity \cite{2004_Paulraj_overviewMIMO}. To meet surging mobile data demands, research and industry have focused on expanding spectrum bandwidth and spatial degrees of freedom. Given spectrum scarcity, deploying large-scale antenna arrays at base stations (BSs) has become essential for fifth-generation and beyond, driving MIMO toward ultra- and extremely large-scale systems with unprecedented beamforming and multiplexing gains \cite{2014_Larsson_massiveMIMO,2024_Zhe_MIMO}. However, scaling up antennas incurs high hardware cost, energy consumption, and processing complexity, underscoring the need for more energy- and cost-efficient solutions \cite{2020_Chowdhury_challenges}.

Recently, intelligent reflecting surfaces (IRSs) \cite{2019_Qingqing_Joint}, also known as reconfigurable intelligent surfaces \cite{2019_Chongwen_RIS}, have emerged as a cost- and energy-efficient technology for enhancing wireless communications. An IRS comprises numerous passive metamaterial elements that can dynamically adjust the amplitude and phase of incident electromagnetic waves, enabling signal enhancement and interference suppression by reconfiguring the propagation environment. Owing to their lightweight structure and flexible deployment, IRSs are suitable for diverse applications \cite{2025_Qingqing_deployment}. When the element count is large, the received signal power or signal-to-noise ratio (SNR) gain scales quadratically with the number of elements \cite{2019_Qingqing_Joint}, motivating extensive research in various systems \cite{2020_Shuowen_Capacity,2019_Miao_Secure,2020_Xinrong_AN,2021_Meng_CoMP,2021_Jianyue_IRS_NOMA,2022_Weidong_IRS,2024_Ying_Grouping}. In particular, IRS-enabled coverage enhancement has been widely studied \cite{2021_Haiquan_coverage,2022_Wenyan_coverage,2023_Guangji_coverage,2023_Jie_coverage_dualIRS,2024_Hanan_coverage}. For instance, \cite{2021_Haiquan_coverage} jointly optimized active and passive beamforming and aerial IRS deployment to maximize worst-case SNR, \cite{2023_Guangji_coverage} proposed a static regulated IRS architecture leveraging distributed MIMO for spatial diversity, and \cite{2023_Jie_coverage_dualIRS} employed two distributed IRSs to improve design flexibility and extend coverage.

However, the above studies are based on fixed-position antenna (FPA) architectures, where antenna locations and orientations are preset during manufacturing. While simplifying production and lowering initial costs, FPAs suffer in dynamic environments, offering limited spatial diversity and reduced performance under deep fading \cite{2023_Lipeng_overview}. Movable antennas (MAs) \cite{2023_Lipeng_Modeling}, also known as fluid antennas \cite{2021_Wong_fluid}, address these limitations by enabling mechanical position adjustment within a defined area to adapt to channel variations. Unlike FPAs subject to static fading, MAs enhance signal quality by relocating to favorable positions, enabling superior performance with the same or fewer antennas and radio frequency (RF) chains. Early works \cite{2023_Lipeng_Modeling,2023_Wenyan_MIMO,2023_Lipeng_uplink,2023_Lipeng_null,2024_Weidong_graph} have shown that MAs significantly improve signal enhancement, interference suppression, beamforming flexibility, and multiplexing, making them a key enabler for future networks. Building on these advantages, MAs have been applied to diverse scenarios, including multiuser uplink/downlink \cite{2024_Nian_MA,2024_Zhenyu_uplink,2024_Ziyuan_twotime}, interference channels \cite{2024_Honghao_IFC}, multicast \cite{2024_Ying_multicast}, secure communications \cite{2024_Guojie_secure_lett,2025_Jun_MAsecure}, cognitive radio \cite{2024_Weidong_MA_spectrumsharing}, and wireless-powered communications \cite{2024_Ying_WPCN}.

Although both IRS and MA technologies offer significant potential for enhancing wireless communications, they differ fundamentally in operating mechanisms and application scenarios. An IRS passively reconfigures the wireless propagation environment by dynamically adjusting phase shifts, yet it cannot transmit independently and relies entirely on existing communication systems. In contrast, MAs actively optimize channel conditions through sub-wavelength mechanical position adjustments, although their flexibility is inherently constrained by mechanical range and physical boundaries. These distinctions highlight their complementary strengths: IRSs can be flexibly deployed on building surfaces or aerial platforms to mitigate coverage blind spots by establishing non-line-of-sight (NLoS) paths, whereas MAs fully exploit channel variations across the continuous spatial domain through dynamic repositioning. Moreover, integrating IRSs enriches multipath diversity, thereby enabling MAs to leverage spatial variations more effectively for performance enhancement \cite{2025_Lipeng_Tutorial}. Recent advances have demonstrated that combining IRSs with MAs yields substantial performance gains over conventional IRS-assisted FPA systems \cite{2025_Qingqing_MAIRS}, including reduced outage probability and delay outage rate \cite{2024_Rostami_fluidRIS}, enhanced throughput \cite{2025_Yunan_MARIS,2024_Junteng_fluidRIS,2024_Weidong_MAIRS,2025_Weidong_MAIRS}, improved physical-layer security \cite{2025_Rostami_fluidRIS_secrecy}, and superior integration of sensing and communication \cite{2025_Haisu_MARIS_ISAC}.

Despite these promising advances, to the best of the authors' knowledge, the potential of integrating MAs and IRSs for coverage enhancement remains unexplored. In scenarios involving multiple target areas, the inherent reconfigurability enabled by both MAs and IRSs gives rise to a variety of coverage strategies with different degrees of freedom. While greater flexibility typically leads to better performance, it also entails higher implementation complexity and energy consumption. This motivates the investigation of diverse MA-IRS integration schemes that achieve a balanced trade-off between performance and cost, thereby offering flexible options for different application requirements. Furthermore, a key design consideration in such integrated systems lies in the allocation of resources, particularly the number of active MAs versus passive IRS elements. Given the significant cost disparity between active MAs (supporting active beamforming) and passive IRS elements (enabling passive beamforming), understanding the trade-off between active and passive components is critical for maximizing system performance under a given total cost.  \looseness=-1

Motivated by the above considerations, this paper studies a multi-IRS-assisted MA system comprising multiple IRSs, a multi-MA BS, and several designated target areas, as illustrated in Fig. \ref{Fig:system_model}. We aim to maximize the worst-case SNR across all the target areas by jointly optimizing the active and passive beamforming along with the MA positioning, subject to constraints on the power-normalized transmit beamforming direction, the finite moving region of each MA, the minimum inter-MA distance, and the modulus of the IRS phase shifts. Our main contributions are summarized as follows: 

\begin{figure}[!t]
	\centering
	\includegraphics[scale=0.67]{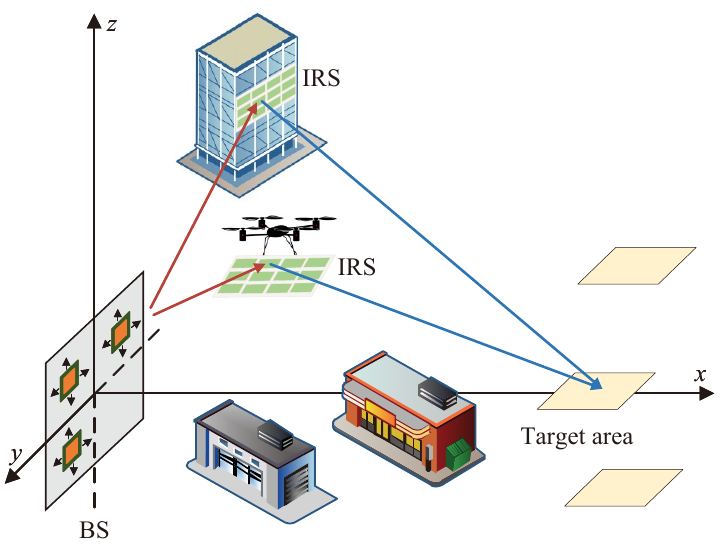}
	\caption{Illustration of integrating MAs and IRSs for coverage enhancement.} \label{Fig:system_model}
	\vspace{-3mm}
\end{figure} 

\begin{figure*}[!t]
	\hspace{-1mm}
	\subfigure[Area-adaptive MA-IRS scheme.]{\label{fig:scheme_a}
		\includegraphics[width = 0.31\textwidth]{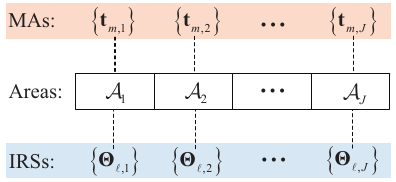}}
	\hspace{1mm}
	\subfigure[Area-adaptive MA-staIRS scheme.]{\label{fig:scheme_b}
		\includegraphics[width = 0.31\textwidth]{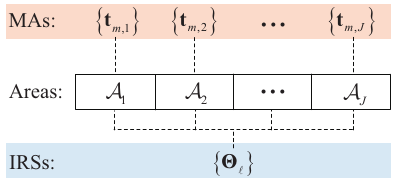}}
	\hspace{1mm}	
	\subfigure[Shared MA-staIRS scheme.]{\label{fig:scheme_c}
		\includegraphics[width = 0.31\textwidth]{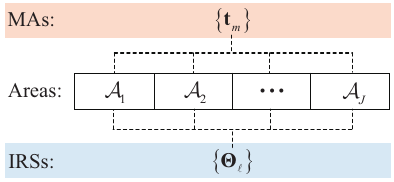}} 
	\caption{Illustration of three different coverage schemes, where the dashed lines indicate the correspondence between MA-IRS solution pairs and their designated target areas.}
	\vspace{-2mm}
\end{figure*}

\begin{itemize}
	\item To balance performance and cost, we propose three coverage-enhancement schemes: the area-adaptive MA-IRS, the area-adaptive MA-staIRS, and the shared MA-staIRS, where staIRS denotes static IRSs with phase shifts configured once during installation. These schemes offer progressively reduced adaptability and cost, from fully adaptive MA positioning with IRS configuration, to MA-only adaptation with static IRSs, and finally to shared MA positioning with static IRSs for all target areas. The corresponding worst-case SNR maximization problems, denoted (P1)-(P3), involve highly coupled variables and implicit objectives, making them challenging to solve optimally.  
	\item To tackle these problems, we develop a general algorithmic framework, illustrated with (P1) as an example. We first obtain the optimal transmit beamforming for given MA positions and IRS phase shifts, and derive an explicit expression for the expected SNR at any target location. The continuous spatial domain is then discretized via uniform sampling to produce a tractable approximation of (P1). Finally, an alternating optimization (AO) algorithm iteratively updates the MA positions and IRS phase shifts until convergence to a suboptimal solution.
	\item Simulation results show that the proposed MA-based designs deliver significant performance gains over their FPA-based counterparts, with the area-adaptive MA-IRS scheme consistently achieving the highest worst-case SNR. Furthermore, systems employing static IRSs with a limited number of MAs may underperform FPA systems with adaptive IRSs, but a modest antenna increase can reverse this while keeping antenna numbers far below IRS elements. Additionally, under a fixed total cost with a linear cost model, the optimal MA-to-IRS-element ratio for maximizing worst-case SNR scales inversely with their unit cost ratio, offering practical guidance for the cost-limited design of IRS-aided MA systems.
\end{itemize}

The subsequent sections of this paper are organized as follows. Section \ref{Sec_model_formu} elaborates on the system model and formulates the optimization problems for a multi-IRS-assisted MA system under three different coverage schemes. Section \ref{Sec_Prob_solution} presents a general algorithmic framework to address these problems. Numerical results are provided in Section \ref{Sec_simulation} to validate the performance of the proposed schemes. Finally, Section \ref{Sec_conclusion} concludes the paper.  

\emph{Notations:} The set of complex numbers is denoted by $\mathbb{C}$, and $\mathbb{C}^{M \times N}$ denotes the space of $M \times N$ complex matrices. We use $(\cdot)^T$, $(\cdot)^{\mathrm H}$, and $\mathbb{E}[\cdot]$ for transpose, Hermitian transpose, and expectation, respectively. For a vector $\mathbf{a}$, ${\rm diag}(\mathbf{a})$ forms a diagonal matrix, while for a matrix $\mathbf{A}$, ${\rm Diag}(\mathbf{A})$ extracts its diagonal into a vector. The cardinality of set $\mathcal{K}$ is $|\mathcal{K}|$. A complex Gaussian distribution with mean $\bm{\mu}$ and covariance $\mathbf{\Sigma}$ is denoted by $\mathcal{CN}(\bm{\mu}, \mathbf{\Sigma})$. The magnitude and real part of $x \in \mathbb{C}$ are $\left|x\right|$ and ${\rm Re}{x}$. For $\mathbf{x} \in \mathbb{C}^{M\times 1}$, $\left\|\mathbf{x}\right\|$ denotes its Euclidean norm and $[\mathbf{x}]_i$ its $i$-th entry. For $\mathbf{X} \in \mathbb{C}^{M \times N}$, $\left\|\mathbf{X}\right\|_2$, $\left\|\mathbf{X}\right\|F$, and $\left[\mathbf{X}\right]_{i,j}$ denote its spectral norm, Frobenius norm, and $(i,j)$-th entry. The notation $\mathbf{S}_1 \succeq \mathbf{S}_2$ means $\mathbf{S}_1 - \mathbf{S}_2$ is positive semidefinite. The imaginary unit is $\jmath$ with $\jmath^2=-1$.

\section{System Model and Problem Formulation}\label{Sec_model_formu}
As shown in Fig. \ref{Fig:system_model}, we consider a communication system where $L$ passive IRSs are deployed to assist a BS equipped with $M$ transmit MAs, extending the BS's communication coverage to $J$ designated target areas.\footnote{Passive IRSs consume only negligible control power and can be flexibly deployed on building facades, rooftops, ceilings, or aerial platforms, providing low-cost and energy-efficient localized coverage enhancement as a complement to existing BSs, unlike deploying additional micro BSs which requires active RF chains, power amplifiers, and high-capacity backhaul, and dedicated site acquisition.} \footnote{The target areas represent coverage hotspots or blockage-prone regions that are known a priori and remain relatively stable over long periods (e.g., factory floors, logistics warehouses, or persistent dead zones). Their quasi-static nature enables a tractable yet practically relevant formulation, allowing systematic optimization of MA and IRS resources for realistic deployments.} 
One of the $J$ target areas is covered each time, either in a round-robin fashion or in a specific order.  
The sets of IRSs, MAs, and target areas are denoted by $\mathcal L$, $\mathcal M$, and $\mathcal J$, respectively, with cardinalities $\left| \mathcal L\right| = L$, $\left|\mathcal M\right| = M$, and $\left|\mathcal J\right|  = J$. The $\ell$-th IRS is equipped with $N_\ell$ reflecting elements, and $\mathcal N$ denotes the set of all IRS elements, where $\left|\mathcal N\right| = N = \sum_{\ell=1}^LN_\ell$. 

Without loss of generality, all locations are defined in a three-dimensional Cartesian coordinate system, with the BS reference point at the origin $\mathbf o = [0,0,0]^T$. The moving region of the transmit MAs, denoted by $\mathcal C$, is a square of size $A \times A$ in the $y$–$z$ plane. For IRS $\ell$, the coordinates of its $N_\ell$ elements are $\mathbf p_{\ell,1},\ldots,\mathbf p_{\ell,N_\ell} \in \mathbb R^{3\times 1}$, where $\mathbf p_{\ell,1}$ is the reference point. The $j$-th target area is $\mathcal A_j \subset \mathbb R^{3\times 1}$, with any location denoted by $\mathbf u_j$. Let $d_\ell = \left\|\mathbf p_{\ell,1}\right\|$ and $d_{\mathbf u_j} = \left\|\mathbf u_j\right\|$ denote the distances from the BS reference point to the $\ell$-th IRS reference point and to location $\mathbf u_j \in \mathcal A_j$, respectively. The distance from the $\ell$-th IRS reference point to $\mathbf u_j$ is given by $d_{\ell,j}(\mathbf u_j) = \left\|\mathbf p_{\ell,1} - \mathbf u_j\right\|$. 
We consider three coverage-enhancement schemes: area-adaptive MA-IRS, area-adaptive MA-staIRS, and shared MA-staIRS.

\vspace{-2mm}
\subsection{Area-adaptive MA-IRS Scheme} 
In this scheme, the MA positions and IRS phase shifts are adaptively adjusted for different target areas. As a result, each target area is served by a dedicated MA-IRS configuration, as shown in Fig. \ref{fig:scheme_a}.\footnote{To synchronize multiple IRSs, a central controller (e.g., the BS) broadcasts configuration commands to all IRSs via dedicated reliable control links (e.g., wireless or optical fiber). Each IRS, equipped with a local controller, stores the received command and executes it in a synchronized manner upon receiving a trigger. This trigger can be realized using time-stamping, an external reference signal, or a dedicated hardware-based mechanism.} For the $j$-th target area, the position of the $m$-th transmit MA is denoted as $\mathbf t_{m,j} = \left[0, y_{m,j}, z_{m,j}\right]^T$. The reflection-coefficient matrix applied at IRS $\ell$ for target area $j$ is given by $\mathbf \Theta_{\ell,j} = {\rm diag}\left(\beta_{\ell,1,j}e^{\jmath \psi_{\ell,1,j}}, \ldots, \beta_{\ell,N_\ell,j}e^{\jmath \psi_{\ell,N_\ell,j}}\right) \in \mathbb C^{N_\ell\times N_\ell}$, where $\beta_{\ell,n,j} \in [0,1]$ and $\psi_{\ell,n,j} \in [0,2\pi)$ represent the amplitude reflection coefficient and phase shift of IRS element $n \in \left\lbrace 1, \ldots, N_\ell\right\rbrace$, respectively. Here, we set $\beta_{\ell,n,j} = 1$, $\forall \ell,n,j$, to maximize the signal reflection. 
The sets $\mathcal T_j \triangleq \left\lbrace\mathbf t_{m,j}\right\rbrace_{m=1}^M$ and $\mathcal R_j \triangleq \left\lbrace \mathbf \Theta_{\ell,j}\right\rbrace_{\ell=1}^L$ are defined for compact notation. Additionally, we assume that the sizes of the MA moving region and each IRS are significantly smaller than the corresponding signal propagation distances. This ensures that the far-field condition holds for all links. Under this assumption, variations in the MA positions affect only the phase component of the complex channel coefficients, while the angle of departure (AoD), angle of arrival (AoA), and amplitude remain unchanged \cite{2023_Lipeng_Modeling}. 

It is assumed that line-of-sight (LoS) paths exist in the IRS-related links (i.e., BS-IRS and IRS-user), which can be practically ensured by deploying IRSs at elevated positions such as building facades, rooftops, ceilings, or aerial platforms. Accordingly, all IRS-related links are modeled by the Rician fading model to capture both the deterministic LoS component and the random scattering effects. For the $j$-th target area, the channel from the BS to IRS $\ell$ can be modeled as 
\begin{align}\label{eq:channel_BS_IRS}
	&\mathbf G_{\ell}(\mathcal T_j) = \sqrt{C_0d_{\rm \ell}^{-\alpha_\ell}}\left(\sqrt{\frac{\kappa_\ell}{\kappa_\ell + 1}}\overline{\mathbf G}_\ell(\mathcal T_j) + \sqrt{\frac{1}{\kappa_\ell + 1}}\tilde{\mathbf G}_\ell(\mathcal T_j) \right)\nonumber\\
	& \in \mathbb C^{N_\ell\times M}, 
\end{align}
where $\kappa_\ell$ is the Rician factor for the BS-to-IRS $\ell$ link, and the term $C_0d_{\rm \ell}^{-\alpha_\ell}$ represents the distance-dependent large-scale path loss, with $C_0$ denoting the reference path loss at a distance of $1$ meter (m), and $\alpha_\ell$ being the corresponding path loss exponent. Moreover, $\overline{\mathbf G}_\ell(\mathcal T_j)$ and $\tilde{\mathbf G}_\ell(\mathcal T_j)$ denote the LoS and NLoS channel components, respectively. The LoS component $\overline{\mathbf G}_\ell(\mathcal T_j)$ is deterministic and expressed as
\begin{align}
	& \overline{\mathbf G}_\ell(\mathcal T_j) \nonumber\\
	& = \!\left[1, e^{\jmath\frac{2\pi}{\lambda}\left( \mathbf p_{\ell,2}-\mathbf p_{\ell,1}\right)^T\mathbf a(\theta^{\rm r}_\ell, \phi^{\rm r}_\ell)}, \ldots, e^{\jmath\frac{2\pi}{\lambda}\left( \mathbf p_{\ell,N_\ell} - \mathbf p_{\ell,1}\right)^T\mathbf a(\theta^{\rm r}_\ell, \phi^{\rm r}_\ell)}\!\right]^{\mathrm H} \nonumber\\
	&\times \left[e^{\jmath\frac{2\pi}{\lambda}\mathbf t_{1,j}^T\mathbf a(\theta^{\rm t}_\ell, \phi^{\rm t}_\ell)}, \ldots, e^{\jmath\frac{2\pi}{\lambda}\mathbf t_{M,j}^T\mathbf a(\theta^{\rm t}_\ell, \phi^{\rm t}_\ell)}\right] \triangleq \mathbf e_\ell \mathbf g^{\mathrm H}_\ell(\mathcal T_j),
\end{align}
where $\mathbf a(\theta^{\rm t}_\ell, \phi^{\rm t}_\ell) \triangleq \left[\cos\theta^{\rm t}_\ell\cos\phi^{\rm t}_\ell,\cos\theta^{\rm t}_\ell\sin\phi^{\rm t}_\ell, \sin\theta^{\rm t}_\ell\right]^T$ and $\mathbf a(\theta^{\rm r}_\ell, \phi^{\rm r}_\ell) \triangleq \left[\cos\theta^{\rm r}_\ell\cos\phi^{\rm r}_\ell, \cos\theta^{\rm r}_\ell\sin\phi^{\rm r}_\ell, \sin\theta^{\rm r}_\ell\right]^T$ represent the normalized transmit and receive direction vectors of the LoS path, respectively \cite{2023_Lipeng_uplink}. The parameters $\theta^{\rm t}_\ell$, $\phi^{\rm t}_\ell$, $\theta^{\rm r}_\ell$, and $\phi^{\rm r}_\ell$ denote the elevation AoD, azimuth AoD, elevation AoA, and azimuth AoA, respectively. In addition, $\mathbf e_\ell\triangleq \left[1, e^{\jmath\frac{2\pi}{\lambda}\left(\mathbf p_{\ell,2}-\mathbf p_{\ell,1}\right)^T\mathbf a(\theta^{\rm r}_\ell, \phi^{\rm r}_\ell)}, \ldots, e^{\jmath\frac{2\pi}{\lambda}\left( \mathbf p_{\ell,N_\ell} - \mathbf p_{\ell,1}\right)^T\mathbf a(\theta^{\rm r}_\ell, \phi^{\rm r}_\ell)}\right]^{\mathrm H} \in\mathbb C^{N_{\ell}\times 1}$ and $\mathbf g^{\mathrm H}_\ell(\mathcal T_j) \triangleq  \left[e^{\jmath\frac{2\pi}{\lambda}\mathbf t_{1,j}^T\mathbf a(\theta^{\rm t}_\ell, \phi^{\rm t}_\ell)}, \ldots, e^{\jmath\frac{2\pi}{\lambda}\mathbf t_{M,j}^T\mathbf a(\theta^{\rm t}_\ell, \phi^{\rm t}_\ell)}\right] \in \mathbb C^{1\times M}$ denote the receive and transmit array response vectors, respectively. 
On the other hand, the NLoS component $\tilde{\mathbf G}_\ell(\mathcal T_j)$ is approximately modeled as an independent and identically distributed (i.i.d.) Rayleigh fading matrix, i.e., $\tilde{\mathbf G}_\ell(\mathcal T_j) \sim \mathcal{CN}(\mathbf 0, \mathbf I)$, under the assumptions that the minimum spacing between the MAs is at least $\lambda/2$, the IRS elements are spaced at least $\lambda/2$ apart, and the propagation environment exhibits rich and uncorrelated scattering. Here, $\lambda$ denotes the carrier wavelength.  

Similarly, the channel from IRS $\ell$ to location $\mathbf u_j$ is modeled as 
\begin{align}
	& \mathbf h_{\ell,j}^{\mathrm H}(\mathbf u_j) = \sqrt{C_0d_{\ell,j}^{-\alpha_{\ell,j}}(\mathbf u_j)} \nonumber\\
	& \times\left(\sqrt{\frac{\kappa_{\ell,j}}{\kappa_{\ell,j} + 1}}\overline{\mathbf h}_{\ell,j}^{\mathrm H}(\mathbf u_j) + \sqrt{\frac{1}{\kappa_{\ell,j} + 1}}\tilde{\mathbf h}_{\ell,j}^{\mathrm H}(\mathbf u_j) \right) \in \mathbb C^{1\times N_\ell}, 
\end{align}
where $\kappa_{\ell,j}$ denotes the Rician factor for the link between IRS $\ell$ and location $\mathbf u_j$, and the term $C_0d_{\ell,j}^{-\alpha_{\ell,j}}(\mathbf u_j)$ stands for the distance-dependent large-scale path loss with $\alpha_{\ell,j}$ representing the corresponding path loss exponent. Furthermore, the deterministic LoS component $\overline{\mathbf h}_{\ell,j}^{\mathrm H}(\mathbf u_j)$ is given by
\begin{align}
	\overline{\mathbf h}_{\ell,j}^{\mathrm H}(\mathbf u_j) = & \left[1, e^{\jmath\frac{2\pi}{\lambda}\left(\mathbf p_{\ell,2} -\mathbf p_{\ell,1}\right)^T\mathbf a\left(\theta^{\rm t}_{\ell,j}(\mathbf u_j),\phi^{\rm t}_{\ell,j}(\mathbf u_j)\right)}, \ldots,\right.  \nonumber\\
	& \left. \quad e^{\jmath\frac{2\pi}{\lambda}\left( \mathbf p_{\ell,N_\ell}-\mathbf p_{\ell,1}\right) ^T\mathbf a\left(\theta^{\rm t}_{\ell,j}(\mathbf u_j),\phi^{\rm t}_{\ell,j}(\mathbf u_j)\right)}\right],
\end{align} 
where $\mathbf a\left(\theta^{\rm t}_{\ell,j}(\mathbf u_j),\phi^{\rm t}_{\ell,j}(\mathbf u_j)\right) \triangleq \Big[\!\cos\theta^{\rm t}_{\ell,j}\!(\mathbf u_j)\!\cos\phi^{\rm t}_{\ell,j}\!(\mathbf u_j),\cos\theta^{\rm t}_{\ell,j}\!(\mathbf u_j)\!\sin\phi^{\rm t}_{\ell,j}\!(\mathbf u_j),\sin\theta^{\rm t}_{\ell,j}\!(\mathbf u_j)\!\Big]^T$ denotes the normalized wave vector of the LoS path from IRS $\ell$ to location $\mathbf u_j$. The parameters $\theta^{\rm t}_{\ell,j}(\mathbf u_j)$ and $\phi^{\rm t}_{\ell,j}(\mathbf u_j)$ denote the elevation and azimuth AoDs, respectively. On the other hand, the elements of the NLoS component $\tilde{\mathbf h}_{\ell,j}^{\mathrm H}(\mathbf u_j)$ are assumed to be i.i.d. complex Gaussian random variables, each following $\mathcal{CN}(0,1)$. 

The LoS path between the BS and each target area is assumed to be blocked by obstacles. In such NLoS environments, which are typically characterized by rich multipath scattering, the direct BS-user links are modeled as Rayleigh fading, with the channel to location $\mathbf u_j$ given by 
\begin{align}
	\mathbf f_{\mathbf u_j} = \sqrt{C_0d_{\mathbf u_j}^{-\alpha_j}}\tilde{\mathbf f}_{\mathbf u_j},
\end{align}
where $C_0d_{\mathbf u_j}^{-\alpha_j}$ denotes the large-scale path loss with exponent $\alpha_j$, and $\tilde{\mathbf f}_{\mathbf u_j}$ denotes the i.i.d. Rayleigh fading vector with entries following $\mathcal{CN}(0,1)$. 

Let $P$ and $\mathbf w_{\mathbf u_j}$ denote the transmit power and beamforming vector at the BS for the user located at ${{\mathbf {u}}_j}$, respectively, with $\left\| \mathbf w_{\mathbf u_j}\right\| = 1$. By ignoring the power of the signals reflected or scattered by the IRSs two or more times, the received SNR at location $\mathbf u_j \in \mathcal A_j, j\in\mathcal J$ can be expressed as 
\begin{align}
	&\gamma_j^{(1)}(\mathcal T_j, \mathcal R_j, \mathbf w_{\mathbf u_j},\mathbf u_j) \nonumber\\
	& \hspace{5mm} = \bar P\left|\left( \sum_{\ell=1}^L\mathbf h_{\ell,j}^{\mathrm H}(\mathbf u_j)\mathbf \Theta_{\ell,j}\mathbf G_\ell(\mathcal T_j) + \mathbf f_{\mathbf u_j}\right) \mathbf w_{\mathbf u_j}\right|^2,
\end{align}
where $ \bar P \triangleq \frac{P}{\sigma^2}$ with $\sigma^2$ being the noise power at location $\mathbf u_j$.  

\subsection{Area-adaptive MA-staIRS Scheme}
This scheme is illustrated in Fig. \ref{fig:scheme_b}, where the MA positions are adaptively adjusted for different areas, while the IRSs maintain the static phase shifts configured during their initial deployment. Accordingly, the received SNR at location $\mathbf u_j \in \mathcal A_j, j\in\mathcal J$ is given by $\gamma_j^{(2)}(\mathcal T_j, \left\lbrace \mathbf \Theta_\ell\right\rbrace, \mathbf w_{\mathbf u_j},\mathbf u_j) = \bar P\left|\left( \sum_{\ell=1}^L\mathbf h_{\ell,j}^{\mathrm H}(\mathbf u_j)\mathbf \Theta_{\ell}\mathbf G_\ell(\mathcal T_j) + \mathbf f_{\mathbf u_j}\right) \mathbf w_{\mathbf u_j}\right|^2$, where $\mathbf \Theta_{\ell} = {\rm diag}\left(a_{\ell,1}, \ldots, a_{\ell,N_\ell}\right)$ denotes the reflection-coefficient matrix related to the $\ell$-th IRS. 

\subsection{Shared MA-staIRS Scheme}
In this scheme, the BS performs a one-time adjustment to the MA positions, while the IRSs maintain the static phase shifts.\footnote{Although the MAs in the shared MA-staIRS scheme are positioned only once, they differ from FPAs in that their structural design retains the ability to be repositioned when needed. This built-in flexibility allows the system to adapt to future changes in coverage needs or environments without physical reinstallation, which would otherwise be required for FPAs.} Consequently, all the target areas share the same MA positions and static IRS phase shifts, as illustrated in Fig. \ref{fig:scheme_c}. Accordingly, the received SNR at location $\mathbf u_j \in \mathcal A_j, j\in\mathcal J$ can be expressed as $\gamma_j^{(3)}(\left\lbrace\mathbf t_m\right\rbrace , \left\lbrace \mathbf \Theta_\ell\right\rbrace , \mathbf w_{\mathbf u_j},\mathbf u_j) = \bar P\left|\left( \sum_{\ell=1}^L\mathbf h_{\ell,j}^{\mathrm H}(\mathbf u_j)\mathbf \Theta_{\ell}\mathbf G_\ell(\mathbf t) + \mathbf f_{\mathbf u_j}\right) \mathbf w_{\mathbf u_j}\right|^2$, where $\mathbf t_m = \left[0, y_m, z_m\right]^T$ denotes the position of the $m$-th MA.

\begin{rem}
	\rm Note that the above three schemes are designed to balance adaptability and implementation cost. They span a solution space from fully adaptive configurations (with per-area MA positioning and IRS phase-shift design) to fully static ones (reusing a common MA and IRS setup across all areas). This structure enables flexible trade-offs among performance, complexity, and signaling overhead, making the schemes applicable to diverse deployment scenarios with varying demands and constraints. 
	
	To further illustrate the applicability of these schemes, typical scenarios for each are described below. The area-adaptive MA-IRS scheme suits performance-sensitive applications with strong coordination capabilities, such as industrial automation, smart hospitals, or emergency systems in large indoor environments. The area-adaptive MA-staIRS scheme fits moderately adaptive settings with limited deployment budgets, including campus networks, public infrastructure, and event-based scenarios. The shared MA-staIRS scheme targets cost- and power-sensitive applications with low adaptability requirements, such as remote monitoring, sensor networks, and rural connectivity.
\end{rem}  

\subsection{Transmission Protocol}
In IRS-assisted MA systems, optimizing MA positions and IRS phase shifts based on instantaneous channel state information (CSI) faces several practical obstacles. First, acquiring accurate CSI through exhaustive scanning of all potential antenna positions is computationally prohibitive and time-consuming. Second, frequent repositioning of antennas to track instantaneous CSI incurs excessive energy consumption. Third, in fast-varying channels, frequent IRS reconfiguration synchronized with BS beamforming is hindered by control latency and limited coordination.

Nevertheless, since the BS, IRSs, and target areas are known and fixed, the LoS components of the involved channels remain stable over relatively long timescales. Moreover, once the MA positions and IRS phase shifts are fixed, estimating the instantaneous CSI of the resulting effective channels becomes far simpler than acquiring the full IRS-related CSI. Motivated by these observations, we adopt a hierarchical transmission protocol: the MA positions and IRS phase shifts are optimized using statistical CSI, while the BS transmit beamforming is designed with instantaneous CSI. 
Here, statistical CSI refers to large-scale and long-term channel characteristics, including path loss, Rician factors, deterministic LoS angles, and the statistical distribution of the NLoS components. In practice, path loss parameters can be obtained from field measurements followed by regression fitting of a path loss model, while Rician factors are estimated from pilot-based measurements averaged over time. The LoS angles and distances are determined directly from the deployment geometry of the BS, IRSs, and target areas. For the NLoS components, their statistical behavior is modeled by an i.i.d. Rayleigh fading distribution. 

\subsection{Problem Formulation}
In this paper, our objective is to maximize the worst-case expected SNR across all the target areas by jointly optimizing the MA positions, the IRS phase shifts, and the transmit beamforming vectors. For the area-adaptive MA-IRS scheme, the problem of interest can be formulated as   
\begin{subequations}\label{P1}
	\begin{eqnarray}
		\hspace{-3.5mm}\text{(P1)}: \hspace{-1mm}&\underset{\substack{\left\lbrace \mathcal T_j\right\rbrace,\\ \left\lbrace \mathcal R_j\right\rbrace }}{\max}& \hspace{-1mm}\underset{\substack{\mathbf u_j \in \mathcal A_j,\\ j\in\mathcal J}}{\min} \hspace{-1mm}\left\lbrace \mathbb E_{\mathcal B}\hspace{-1mm}\left[\underset{\mathbf w_{\mathbf u_j}}{\max} \hspace{1mm} \gamma_j^{(1)}\left(\mathcal T_j, \mathcal R_j, \mathbf w_{\mathbf u_j},\mathbf u_j\right)\right] \!\right\rbrace \\
		&\hspace{-2cm}\text{s.t.}& \hspace{-1.3cm} \left\|\mathbf w_{\mathbf u_j}\right\| = 1, \ \forall j \in\mathcal J, \label{P1_cons:b}\\
		&& \hspace{-1.3cm} \mathbf t_{m,j} \in\mathcal C, \ \forall m \in\mathcal M, j \in\mathcal J, \label{P1_cons:c}\\
		&& \hspace{-1.3cm} \left\|\mathbf t_{m,j} - \mathbf t_{q,j}\right\|^2 \geq D^2, \ \forall m,q\in\mathcal M, m\neq q, j\in\mathcal J, \label{P1_cons:d} \\
		&& \hspace{-1.3cm} \left|\left[ \mathbf \Theta_{\ell,j}\right]_{n,n} \right| = 1, \ \forall  \ell\in\mathcal L, j\in\mathcal J, n\in\{1,\ldots,N_\ell\}, \label{P1_cons:e} 
	\end{eqnarray}
\end{subequations}
where $\mathcal B \triangleq \left\lbrace \left\lbrace \tilde{\mathbf G}_\ell\right\rbrace_{\ell=1}^L, \left\lbrace \tilde{\mathbf h}_{\ell,j}^{\mathrm H}\right\rbrace_{\ell=1}^L, \tilde{\mathbf f}_{\mathbf u_j}\right\rbrace $, $D$ denotes the minimum distance between any two MAs, and $D \geq \lambda/2$. Similarly, for the area-adaptive MA-staIRS and shared MA-staIRS schemes, the respective worst-case SNR maximization problems can be formulated as follows:
\begin{subequations}\label{P2}
	\begin{eqnarray}
		\hspace{-5mm}\text{(P2)}: \hspace{-3.2mm}&\underset{\substack{\left\lbrace \mathcal T_j\right\rbrace, \\ \left\lbrace \mathbf \Theta_\ell\right\rbrace }}{\max}& \hspace{-3.3mm}\min_{\substack{\mathbf u_j \in \mathcal A_j,\\ j\in\mathcal J}}\left\lbrace \mathbb E_{\mathcal B}\!\left[\underset{\mathbf w_{\mathbf u_j}}{\max} \hspace{1mm} \gamma_j^{(2)}\left(\mathcal T_j, \left\lbrace \mathbf \Theta_\ell\right\rbrace, \mathbf w_{\mathbf u_j},\mathbf u_j\right)\right] \right\rbrace \\ 
		&\text{s.t.}& \hspace{-1mm} \eqref{P1_cons:b} - \eqref{P1_cons:d},\\
		&& \hspace{-1mm} \left|\left[ \mathbf \Theta_\ell\right]_{n,n} \right| = 1, \ \forall  \ell\in\mathcal L,  n\in\{1,\ldots,N_\ell\}, \label{P2_cons:c} 
	\end{eqnarray}
\end{subequations}
\begin{subequations}\label{P3}
	\begin{eqnarray}
		\hspace{-4.8mm}\text{(P3)}: \hspace{-3mm}&\underset{\substack{\left\lbrace \mathbf t_m\right\rbrace, \\ \left\lbrace \mathbf \Theta_\ell\right\rbrace} }{\max}& \hspace{-3.5mm}\min_{\substack{\mathbf u_j \in \mathcal A_j,\\ j\in\mathcal J}} \hspace{-1mm}\left\lbrace \! \mathbb E_{\mathcal B}\!\left[\underset{\mathbf w_{\mathbf u_j}}{\max} \gamma_j^{(3)}(\left\lbrace \mathbf t_m\right\rbrace\!, \left\lbrace \mathbf \Theta_\ell\right\rbrace\!, \mathbf w_{\mathbf u_j},\mathbf u_j)\right] \!\right\rbrace \\ 
		&\text{s.t.}& \hspace{-2mm} \eqref{P1_cons:b}, \eqref{P2_cons:c},\\
		&& \hspace{-2mm} \mathbf t_m \in\mathcal C, \ \forall m \in\mathcal M, \label{P3_cons:c}\\
		&& \hspace{-2mm} \left\|\mathbf t_m - \mathbf t_q\right\|^2 \geq D^2, \ \forall m,q\in\mathcal M, m\neq q. \label{P3_cons:d}
	\end{eqnarray}
\end{subequations}
Problems (P1)-(P3) are all challenging to solve for the following reasons: 1) the expectations over $\mathcal B$ lack explicit expressions; 2) determining the worst-case SNR over $J$ two-dimensional areas is non-trivial due to the continuous nature of the spatial domain in each area; 3) the optimization variables are intricately coupled in the objective functions; 4) constraints \eqref{P1_cons:b}, \eqref{P1_cons:d}, \eqref{P1_cons:e}, \eqref{P2_cons:c}, and \eqref{P3_cons:d} exhibit non-convexity. As a result, (P1)-(P3) are all non-convex optimization problems, making it difficult, if not impossible, to obtain their optimal solutions. Nevertheless, we propose a general algorithmic framework capable of solving these problems suboptimally, with a detailed implementation presented in the following section. \looseness=-1

\section{Proposed general algorithmic framework for Problems (P1)-(P3)}\label{Sec_Prob_solution}
\subsection{Proposed Solution for (P1)}
To solve (P1), we first optimize the transmit beamforming vectors $\left\lbrace\mathbf w_{\mathbf u_j} \right\rbrace$. For any given MA positions $\left\lbrace\mathcal T_j\right\rbrace$ and IRS reflection-coefficient matrices $\left\lbrace\mathcal R_j\right\rbrace$, the optimal beamforming solution $\mathbf w_{\mathbf u_j}^*$ follows the maximum ratio transmission, which can be expressed as a function of $\left\lbrace \mathcal T_j, \mathcal R_j\right\rbrace $, i.e.,  
\begin{align}\label{equ:beam_solu}
	\mathbf w_{\mathbf u_j}^* = \frac{\left( \sum_{\ell=1}^L\mathbf h_{\ell,j}^{\mathrm H}(\mathbf u_j)\mathbf \Theta_{\ell,j}\mathbf G_\ell(\mathcal T_j) + \mathbf f_{\mathbf u_j}\right)^{\mathrm H}}{\left\| \sum_{\ell=1}^L\mathbf h_{\ell,j}^{\mathrm H}(\mathbf u_j)\mathbf \Theta_{\ell,j}\mathbf G_\ell(\mathcal T_j) + \mathbf f_{\mathbf u_j}\right\|}.
\end{align}
Then, we have the following proposition. 
\begin{prop}\label{prop1}
	The expectation of $\gamma_j^{(1)}\left(\mathcal T_j, \mathcal R_j, \mathbf w_{\mathbf u_j}^*,\mathbf u_j\right)$ over $\mathcal B$ can be explicitly expressed as
	\begin{align}
		&\hspace{-1.2mm}\mathbb E_{\mathcal B} \!\left[\gamma_j^{(1)} \! \left(\mathcal T_j, \mathcal R_j, \mathbf w_{\mathbf u_j}^*,\mathbf u_j\right)\!\right] \!=\! \bar P\left\|\sum_{\ell=1}^L\hat{\mathbf h}_{\ell,j}^{\mathrm H}(\mathbf u_j)\mathbf \Theta_{\ell,j}\hat{\mathbf G}_\ell(\mathcal T_j)\right\|^2 \nonumber\\
		& + \bar P\left( \sum_{\ell=1}^L\left[\left(\kappa_{\ell,j} + \kappa_\ell + 1\right) \Lambda_{\mathbf u_j}MN_{\ell}\right] + C_0d_{\mathbf u_j}^{-\alpha_j}M\right), 
	\end{align}
	where $\hat{\mathbf h}^{\mathrm H}_{\ell,j}(\mathbf u_j)\triangleq \sqrt{C_0d_{\ell,j}^{-\alpha_{\ell,j}}(\mathbf u_j)}\sqrt{\frac{\kappa_{\ell,j}}{\kappa_{\ell,j} + 1}}\overline{\mathbf h}_{\ell,j}^{\mathrm H}(\mathbf u_j)$,  $\hat{\mathbf G}_{\ell}(\mathcal T_j) \triangleq  \sqrt{C_0d_{\ell}^{-\alpha_{\ell}}}\sqrt{\frac{\kappa_{\ell}}{\kappa_{\ell} + 1}}\overline{\mathbf G}_{\ell}^{\mathrm H}(\mathcal T_j)$, and $\Lambda_{\mathbf u_j} \triangleq \frac{C_0^2d_{\ell,j}^{-\alpha_{\ell,j}}(\mathbf u_j)d_{\rm \ell}^{-\alpha_\ell}}{\left(\kappa_{\ell,j}+1\right)\left(\kappa_\ell+1\right)}$.
\end{prop}
\begin{proof}
Please refer to the Appendix. 
\end{proof}	

By applying Proposition \ref{prop1}, problem (P1) can be equivalently transformed into
\begin{subequations}\label{P1_eqv}
	\begin{eqnarray}
		&\hspace{-9mm}\underset{\substack{\{\mathcal T_j\},\\ \{\mathcal R_j\}}}{\max}& \hspace{-4mm} \min_{\substack{\mathbf u_j \in \mathcal A_j,\\ j\in\mathcal J}}\left\lbrace \bar P\left\|\sum_{\ell=1}^L\hat{\mathbf h}_{\ell,j}^{\mathrm H}(\mathbf u_j)\mathbf \Theta_{\ell,j}\hat{\mathbf G}_\ell(\mathcal T_j)\right\|^2 + C_{\mathbf u_j}\right\rbrace \\ 
		&\hspace{-3mm}\text{s.t.}&  \eqref{P1_cons:c}-\eqref{P1_cons:e},
	\end{eqnarray}
\end{subequations}
where $C_{\mathbf u_j} \triangleq \bar P\big( \sum_{\ell=1}^L\left[\left(\kappa_{\ell,j} + \kappa_\ell + 1\right)\Lambda_{\mathbf u_j}MN_{\ell}\right] + C_0d_{\mathbf u_j}^{-\alpha_j}M\big)$. Although problem \eqref{P1_eqv} addresses the challenge of non-closed-form expectation and involves fewer optimization variables compared to the original problem (P1), it remains difficult to solve since each $\mathbf u_j$ can take values from an uncountably infinite set $\mathcal A_j$. To overcome this, we uniformly sample each target area $\mathcal A_j$ to generate a countable set of candidate locations, denoted by $\mathcal G_j$. By replacing $\mathcal A_j$ with $\mathcal G_j$ and defining $\eta \triangleq \min_{\mathbf u_j \in \mathcal G_j, j\in\mathcal J} \left\lbrace\bar P\left\|\sum_{\ell=1}^L\hat{\mathbf h}_{\ell,j}^{\mathrm H}(\mathbf u_j)\mathbf \Theta_{\ell,j}\hat{\mathbf G}_\ell(\mathcal T_j)\right\|^2 + C_{\mathbf u_j}\right\rbrace$, problem \eqref{P1_eqv} becomes 
\begin{subequations}\label{P1_eqv_appro}
	\begin{eqnarray}
		&\underset{\eta, \{\mathcal T_j\}, \{\mathbf \Theta_{j}\}}{\max}& \hspace{-2mm} \eta \\
		&\text{s.t.}& \hspace{-7mm} \bar P\left\|\sum_{\ell=1}^L\hat{\mathbf h}_{\ell,j}^{\mathrm H}(\mathbf u_j)\mathbf \Theta_{\ell,j}\hat{\mathbf G}_\ell(\mathcal T_j)\right\|^2 + C_{\mathbf u_j} \geq \eta, \nonumber\\
		&& \hspace{-7mm} \forall \mathbf u_j \in \mathcal G_j, j\in\mathcal J, \label{P1_eqv_appro_cons:b}\\ 
		&& \hspace{-7mm} \eqref{P1_cons:c}-\eqref{P1_cons:e}.
	\end{eqnarray}
\end{subequations}
Problem \eqref{P1_eqv_appro} serves as an inner approximation of problem \eqref{P1_eqv}. As the sampling density increases and $\mathcal G_j \rightarrow \mathcal A_j, \forall j\in\mathcal J$, problem \eqref{P1_eqv_appro} becomes asymptotically equivalent to problem \eqref{P1_eqv}. However, a higher sampling density typically results in a larger number of candidate locations, which increases the problem size and prolongs the solving time. Therefore, the sampling density should be carefully controlled to balance computational complexity and performance degradation. In this work, since the MA positions $\{\mathcal T_j\}$ and IRS phase shifts $\{\mathcal R_j\}$ are optimized based on statistical CSI, the design can be conducted offline, which allows for the adoption of a relatively dense sampling strategy without incurring excessive real-time computational burden. 

Next, we focus on solving problem \eqref{P1_eqv_appro}. The non-convex unit-modulus constraints in \eqref{P1_cons:e} are first relaxed to 
\begin{align}
\left|\left[ \mathbf \Theta_{\ell,j}\right]_{n,n} \right| \leq 1, \ \forall  \ell\in\mathcal L, j\in\mathcal J, n\in\{1,\ldots,N_\ell\}, \label{cons:modulus_slack}
\end{align} 
which results in a convex constraint set. This relaxation leads to an upper bound of the optimal value of problem \eqref{P1_eqv_appro}, which can be obtained by solving
\begin{subequations}\label{P1_eqv_appro_slack}
	\begin{eqnarray}
		&\underset{\eta, \{\mathcal T_j\}, \{\mathbf \Theta_{j}\}}{\max}& \eta \\
		&\text{s.t.}& \hspace{-1.5mm} \eqref{cons:modulus_slack}, \eqref{P1_eqv_appro_cons:b},\eqref{P1_cons:c},\eqref{P1_cons:d}. 
	\end{eqnarray}
\end{subequations}
The main difficulties in this problem arise from the severe variable coupling in constraint \eqref{P1_eqv_appro_cons:b} and the inherent non-convexity of constraint \eqref{P1_cons:d}. To overcome these issues, we employ an AO framework to decouple the optimization of $\{\mathcal T_j\}$ and $\{\mathcal R_j\}$, which are updated alternately until convergence, as detailed below. 

\subsubsection{Optimizing $\{\mathcal T_j\}$} For any given $\{\mathcal R_j\}$, the subproblem of optimizing $\{\mathcal T_j\}$ is given by
\begin{subequations}\label{P1_eqv_appro_sub1}
	\begin{eqnarray}
		&\underset{\eta, \{\mathbf \Theta_{j}\}}{\max}& \eta \\
		&\text{s.t.}& \hspace{-1.5mm} \eqref{P1_eqv_appro_cons:b},\eqref{P1_cons:c},\eqref{P1_cons:d}.
	\end{eqnarray}
\end{subequations}
To solve this problem, we first expand the term $\left\|\sum_{\ell=1}^L\hat{\mathbf h}_{\ell,j}^{\mathrm H}(\mathbf u_j)\mathbf \Theta_{\ell,j}\hat{\mathbf G}_\ell(\mathcal T_j)\right\|^2$ in constraint \eqref{P1_eqv_appro_cons:b}, so that the optimization variables $\mathcal T_j \triangleq \{\mathbf t_{m,j}\}$ appear explicitly in the resulting expression. This step is crucial for characterizing the curvature of constraint \eqref{P1_eqv_appro_cons:b} and for developing a tractable solution method. To simplify the notation for the subsequent expansion, we define 
\begin{align}
&\varpi_{\ell,\ell',j}(\mathbf u_j)\triangleq  C_0^2\sqrt{d_{\ell,j}^{-\alpha_{\ell,j}}(\mathbf u_j)d_{\ell',j}^{-\alpha_{\ell',j}}(\mathbf u_j)d_{\ell}^{-\alpha_\ell}d_{ \ell'}^{-\alpha_{\ell'}}} \nonumber\\
& \hspace{1cm} \times\sqrt{\frac{\kappa_{\ell,j}\kappa_{\ell',j}\kappa_\ell\kappa_{\ell'}}{\left( \kappa_{\ell,j} + 1\right) \left(\kappa_{\ell',j} + 1 \right)\left( \kappa_\ell + 1\right) \left(\kappa_{\ell'} + 1 \right)}}.
\end{align}
Then, we obtain the expanded expression in \eqref{equ:term_expansion}, which is shown at the top of the next page,  
\begin{figure*}[!ht]
\begin{align}\label{equ:term_expansion}
  	\hspace{-2mm}\left\|\sum_{\ell=1}^L\hat{\mathbf h}_{\ell,j}^{\mathrm H}(\mathbf u_j)\mathbf \Theta_{\ell,j}\hat{\mathbf G}_\ell(\mathcal T_j)\right\|^2 & = \sum_{\ell=1}^L \sum_{\ell'=1}^L \left[ \varpi_{\ell,\ell',j}(\mathbf u_j)\overline{\mathbf h}_{\ell,j}^{\mathrm H}(\mathbf u_j)\mathbf \Theta_{\ell,j}\overline{\mathbf G}_\ell(\mathcal T_j) \overline{\mathbf G}_{\ell'}^{\mathrm H}(\mathcal T_j)\mathbf\Theta_{\ell',j}^{\mathrm H}\overline{\mathbf h}_{\ell',j}(\mathbf u_j)\right] \nonumber\\
  	& = \sum_{\ell=1}^L\sum_{\ell'=1}^L\left[\varpi_{\ell,\ell',j}(\mathbf u_j)\overline{\mathbf h}_{\ell,j}^{\mathrm H}(\mathbf u_j)\mathbf \Theta_{\ell,j}\mathbf e_\ell\mathbf g^{\mathrm H}_\ell(\mathcal T_j) \mathbf g_{\ell'}(\mathcal T_j)\mathbf e^{\mathrm H}_{\ell'}\mathbf\Theta_{\ell',j}^{\mathrm H}\overline{\mathbf h}_{\ell',j}(\mathbf u_j)\right]\nonumber\\
  	& \triangleq \sum_{\ell=1}^L\sum_{\ell'=1}^L\left[ b_{\ell,\ell',j}(\mathbf u_j)\mathbf g^{\mathrm H}_\ell(\mathcal T_j)\mathbf g_{\ell'}(\mathcal T_j)\right]  = \sum_{\ell=1}^L\sum_{\ell'=1}^L \left[ b_{\ell,\ell',j}(\mathbf u_j)\sum_{m=1}^Me^{\jmath\frac{2\pi}{\lambda}\mathbf t_{m,j}^T\left(\mathbf a\left(\theta^{\rm t}_{\ell},\phi^{\rm t}_{\ell}\right) - \mathbf a\left(\theta^{\rm t}_{\ell'},\phi^{\rm t}_{\ell'}\right) \right)}\right] \nonumber\\
  	& \triangleq \sum_{\ell=1}^L\sum_{\ell'=1}^L\left[ b_{\ell,\ell',j}(\mathbf u_j)\sum_{m=1}^Me^{\jmath\nu_{\ell,\ell'}(\mathbf t_{m,j})}\right]  = \sum_{\ell=1}^L\sum_{\ell'=1}^L\left[ {\rm Re}\left\lbrace b_{\ell,\ell',j}(\mathbf u_j)\right\rbrace \sum_{m=1}^M\cos\left(\nu_{\ell,\ell'}(\mathbf t_{m,j})\right)\right] \nonumber\\
  	& \triangleq \sum_{\ell=1}^L\sum_{\ell'=1}^L\left[ {\rm Re}\left\lbrace b_{\ell,\ell',j}(\mathbf u_j)\right\rbrace \sum_{m=1}^MZ_{\ell,\ell'}(\mathbf t_{m,j})\right],
\end{align}
	\hrulefill
\end{figure*}
where 
\begin{align}
&b_{\ell,\ell',j}(\mathbf u_j) \triangleq \varpi_{\ell,\ell',j}(\mathbf u_j)\mathbf e^{\mathrm H}_{\ell'}\mathbf\Theta_{\ell',j}^{\mathrm H}\overline{\mathbf h}_{\ell',j}(\mathbf u_j)\overline{\mathbf h}_{\ell,j}^{\mathrm H}(\mathbf u_j)\mathbf \Theta_{\ell,j}\mathbf e_\ell \nonumber\\
& \in \mathbb C,
\end{align}
$\nu_{\ell,\ell'}(\mathbf t_{m,j}) \triangleq \frac{2\pi}{\lambda}\big[y_{m,j}^{\rm t}\left(\cos\theta^{\rm t}_\ell\sin\phi^{\rm t}_\ell - \cos\theta^{\rm t}_{\ell'}\sin\phi^{\rm t}_{\ell'}\right) + z_{m,j}^{\rm t}\left(\sin\theta^{\rm t}_\ell - \sin\theta^{\rm t}_{\ell'}\right) \big]$, and $Z_{\ell,\ell'}(\mathbf t_{m,j}) \triangleq \cos\left(\nu_{\ell,\ell'}(\mathbf t_{m,j})\right)$. In \eqref{equ:term_expansion}, the complex coefficients and exponential terms are represented by their real-valued components, namely the real part and cosine function, since the squared norm is real by definition and the imaginary parts cancel out due to conjugate symmetry. Since $Z_{\ell,\ell'}(\mathbf t_{m,j})$ is neither concave nor convex with respect to $\mathbf t_{m,j}$, constraint \eqref{P1_eqv_appro_cons:b} is non-convex. 

Nevertheless, it can be observed that $Z_{\ell,\ell'}(\mathbf t_{m,j})$ exhibits bounded curvature, meaning that there exists a positive constant $\psi_{\ell,\ell',m,j}$ such that $\psi_{\ell,\ell',m,j}\mathbf I \succeq \nabla^2Z_{\ell,\ell'}(\mathbf t_{m,j})$. This motivates the construction of a surrogate function that locally approximates the term $Z_{\ell,\ell'}(\mathbf t_{m,j})$ via a second-order Taylor expansion, so as to obtain a convex approximation of constraint \eqref{P1_eqv_appro_cons:b} and enable the application of the SCA technique. It is worth highlighting that the choice between a lower-bound and an upper-bound surrogate function depends on whether ${\rm Re}\left\lbrace b_{\ell,\ell',j}(\mathbf u_j)\right\rbrace$ is positive or negative, in order to ensure a valid and conservative approximation of the original constraint. Specifically, let $Z_{\ell,\ell'}^{\rm lb,\it r}(\mathbf t_{m,j})$ and $Z_{\ell,\ell'}^{\rm ub,\it r}(\mathbf t_{m,j})$ denote the lower-bound and upper-bound surrogate functions of $Z_{\ell,\ell'}(\mathbf t_{m,j})$ in the $r$-th iteration, respectively. Then, we define
\begin{align}
	\hat Z_{\ell,\ell'}^r(\mathbf t_{m,j}) \triangleq 
	\begin{cases} 
		Z_{\ell,\ell'}^{\rm lb,\it r}(\mathbf t_{m,j}), & \text{if } {\rm Re}\left\lbrace b_{\ell,\ell',j}(\mathbf u_j)\right\rbrace \geq 0, \\
		Z_{\ell,\ell'}^{\rm ub,\it r}(\mathbf t_{m,j}), & \text{if } {\rm Re}\left\lbrace b_{\ell,\ell',j}(\mathbf u_j)\right\rbrace < 0. 
	\end{cases} 
\end{align}
Based on this definition, the following inequality holds: 
\begin{align}
	&\sum_{\ell=1}^L\sum_{\ell'=1}^L\left[ {\rm Re}\left\lbrace b_{\ell,\ell',j}(\mathbf u_j)\right\rbrace \sum_{m=1}^MZ_{\ell,\ell'}(\mathbf t_{m,j})\right]  \nonumber\\
	& \hspace{8mm}\geq \sum_{\ell=1}^L\sum_{\ell'=1}^L\left[ {\rm Re}\left\lbrace b_{\ell,\ell',j}(\mathbf u_j)\right\rbrace \sum_{m=1}^M\hat Z_{\ell,\ell'}^r(\mathbf t_{m,j})\right].
\end{align}
Consequently, a convex approximation of constraint \eqref{P1_eqv_appro_cons:b} is given by 
\begin{align}\label{P1_eqv_appro_cons:b_sca}
&\bar P\sum_{\ell=1}^L\sum_{\ell'=1}^L\left[ {\rm Re}\left\lbrace b_{\ell,\ell',j}(\mathbf u_j)\right\rbrace \sum_{m=1}^M\hat Z_{\ell,\ell'}^r(\mathbf t_{m,j})\right]  + C_{\mathbf u_j} \geq \eta, \nonumber\\
& \forall \mathbf u_j \in \mathcal G_j, j\in\mathcal J.
\end{align}

We next describe in detail how the lower-bound and upper-bound surrogate functions, $Z_{\ell,\ell'}^{\rm lb,\it r}(\mathbf t_{m,j})$ and $Z_{\ell,\ell'}^{\rm ub,\it r}(\mathbf t_{m,j})$, are constructed. According to \cite[Lemma 12]{2017_Sun_MM}, with given local points $\{\mathbf t_{m,j}^r\}$ in the $r$-th iteration, we have
\begin{align}\label{ineq:surrogate_func_omega}
	Z_{\ell,\ell'}(\mathbf t_{m,j}) & \geq  Z_{\ell,\ell'}^{\rm lb,\it r}(\mathbf t_{m,j}) \triangleq Z_{\ell,\ell'}(\mathbf t_{m,j}^r) + \nabla Z_{\ell,\ell'}(\mathbf t_{m,j}^r)^T \nonumber\\
	& \hspace{4mm}\times (\mathbf t_{m,j} - \mathbf t_{m,j}^r) 
	 - \frac{\psi_{\ell,\ell',m,j}}{2}(\mathbf t_{m,j} - \mathbf t_{m,j}^r)^T \nonumber\\
	& \hspace{4mm} \times (\mathbf t_{m,j} - \mathbf t_{m,j}^r) , 
\end{align}
\begin{align}\label{ineq:surrogate_func_omega_2}
	Z_{\ell,\ell'}(\mathbf t_{m,j}) & \leq Z_{\ell,\ell'}^{\rm ub,\it r}(\mathbf t_{m,j}) \triangleq  Z_{\ell,\ell'}(\mathbf t_{m,j}^r) + \nabla Z_{\ell,\ell'}(\mathbf t_{m,j}^r)^T \nonumber\\
	& \hspace{4mm} \times (\mathbf t_{m,j} - \mathbf t_{m,j}^r) + \frac{\psi_{\ell,\ell',m,j}}{2}(\mathbf t_{m,j} - \mathbf t_{m,j}^r)^T \nonumber\\
	& \hspace{4mm} \times (\mathbf t_{m,j} - \mathbf t_{m,j}^r).  
\end{align}
Here, the gradient $\nabla Z_{\ell,\ell'}(\mathbf t_{m,j}^r)^T$ takes the following explicit form:
\begin{align}
	& \nabla Z_{\ell,\ell'}(\mathbf t_{m,j}^r)^T \nonumber\\
	& = \left[\frac{\partial Z_{\ell,\ell'}(\mathbf t_{m,j})}{\partial x_{m,j}^{\rm t}}\Big|_{\mathbf t_{m,j} =\mathbf t_{m,j}^r}, \frac{\partial Z_{\ell,\ell'}(\mathbf t_{m,j})}{\partial y_{m,j}^{\rm t}}\Big|_{\mathbf t_{m,j} = \mathbf t_{m,j}^r}\right],
\end{align}
with 
\begin{subequations}\label{eq:first-order}
	\begin{align}
		&\frac{\partial Z_{\ell,\ell'}(\mathbf t_{m,j})}{\partial y_{m,j}^{\rm t}}\Big|_{\mathbf t_{m,j} = \mathbf t_{m,j}^r} =  -\frac{2\pi}{\lambda} \alpha_{\ell,\ell'}\sin\left(\nu_{\ell,\ell'}(\mathbf t_{m,j}^r)\right), \\
		&\frac{\partial Z_{\ell,\ell'}(\mathbf t_{m,j})}{\partial z_{m,j}^{\rm t}}\Big|_{\mathbf t_{m,j} = \mathbf t_{m,j}^r} = -\frac{2\pi}{\lambda}\omega_{\ell,\ell'}\sin\left(\nu_{\ell,\ell'}(\mathbf t_{m,j}^r)\right).
	\end{align}
\end{subequations}
In \eqref{eq:first-order}, $\alpha_{\ell,\ell'} \triangleq \cos\theta^{\rm t}_\ell\sin\phi^{\rm t}_\ell - \cos\theta^{\rm t}_{\ell'}\sin\phi^{\rm t}_{\ell'}$ and $\omega_{\ell,\ell'} \triangleq \sin\theta^{\rm t}_\ell - \sin\theta^{\rm t}_{\ell'}$. In addition, a conservative yet feasible choice of $\psi_{\ell,\ell',m,j}$ that guarantees $\psi_{\ell,\ell',m,j}\mathbf I \succeq \nabla^2 Z_{\ell,\ell'}(\boldsymbol t_{m,j})$ is to set $\psi_{\ell,\ell',m,j} \geq \left\|\nabla^2 Z_{\ell,\ell'}(\mathbf t_{m,j})\right\|_F$. This selection is justified by the fundamental matrix inequality $\left\|\nabla^2 Z_{\ell,\ell'}(\mathbf t_{m,j})\right\|_F\mathbf I \succeq \left\|\nabla^2 Z_{\ell,\ell'}(\mathbf t_{m,j})\right\|_2\mathbf I \succeq \nabla^2 Z_{\ell,\ell'}(\mathbf t_{m,j})$, which ensures a mathematically sound and conservative upper bound. To continue, the Frobenius norm $\left\|\nabla^2 Z_{\ell,\ell'}(\mathbf t_{m,j})\right\|_F$ can be explicitly computed as follows: 
\begin{align}
	& \left\|\nabla^2 Z_{\ell,\ell'}(\mathbf t_{m,j})\right\|_F  \!=\! \Bigg[\left(\frac{\partial Z_{\ell,\ell'}(\mathbf t_{m,j})}{\partial y_{m,j}^{\rm t}\partial y_{m,j}^{\rm t}}\right)^2 \!+\! \left(\frac{\partial Z_{\ell,\ell'}(\mathbf t_{m,j})}{\partial y_{m,j}^{\rm t}\partial z_{m,j}^{\rm t}}\right)^2 \nonumber\\
	& \hspace{1cm} + \left(\frac{\partial Z_{\ell,\ell'}(\mathbf t_{m,j})}{\partial z_{m,j}^{\rm t}\partial y_{m,j}^{\rm t}}\right)^2  + \left(\frac{\partial Z_{\ell,\ell'}(\mathbf t_{m,j})}{\partial z_{m,j}^{\rm t}\partial z_{m,j}^{\rm t}}\right)^2\Bigg]^{\frac{1}{2}},
\end{align}
where 
\begin{subequations}\label{eq:second-order}
	\begin{align}
		& \frac{\partial Z_{\ell,\ell'}(\mathbf t_{m,j})}{\partial y_{m,j}^{\rm t}\partial y_{m,j}^{\rm t}} = -\frac{4\pi^2}{\lambda^2} \alpha_{\ell,\ell'}^2\cos\left(\nu_{\ell,\ell'}(\mathbf t_{m,j})\right), \\
		& \frac{\partial Z_{\ell,\ell'}(\mathbf t_{m,j})}{\partial y_{m,j}^{\rm t}\partial z_{m,j}^{\rm t}} = \frac{\partial Z_{\ell,\ell'}(\mathbf t_{m,j})}{\partial z_{m,j}^{\rm t}\partial y_{m,j}^{\rm t}} \nonumber\\
		& \hspace{1.9cm} = -\frac{4\pi^2}{\lambda^2} \alpha_{\ell,\ell'}\omega_{\ell,\ell'}\cos\left(\nu_{\ell,\ell'}(\mathbf t_{m,j})\right), \\
		& \frac{\partial Z_{\ell,\ell'}(\mathbf t_{m,j})}{\partial z_{m,j}^{\rm t}\partial z_{m,j}^{\rm t}} =  -\frac{4\pi^2}{\lambda^2} \omega_{\ell,\ell'}^2\cos\left(\nu_{\ell,\ell'}(\mathbf t_{m,j})\right).
	\end{align}
\end{subequations}
By setting $\cos\left(\nu_{\ell,\ell'}(\mathbf t_{m,j})\right) =1$, an upper bound of $\left\|\nabla^2 Z_{\ell,\ell'}(\mathbf t_{m,j})\right\|_F$ can be derived,  which in turn can be used as a valid choice for $\psi_{\ell,\ell',m,j}$. The detailed procedures for addressing the non-convex constraint \eqref{P1_eqv_appro_cons:b} have now been fully presented. 

The only remaining obstacle to solving problem \eqref{P1_eqv_appro_sub1} lies in the non-convex constraint \eqref{P1_cons:d}. Since the function $\left\|\mathbf t_{m,j} - \mathbf t_{q,j}\right\|^2$ is jointly convex in $(\mathbf t_{m,j},\mathbf t_{q,j})$, it admits a global lower bound given by its first-order Taylor expansion at any point. Accordingly, for a given local point $(\mathbf t_{m,j}^r,\mathbf t_{q,j}^r)$ in the $r$-th iteration, we have
\begin{align}
	\left\|\mathbf t_{m,j} - \mathbf t_{q,j}\right\|^2 \geq &  - \left\|\mathbf t_{m,j}^r - \mathbf t_{q,j}^r\right\|^2 + 2\left(\mathbf t_{m,j}^r - \mathbf t_{q,j}^r\right)^T \nonumber\\
	& \times \left(\mathbf t_{m,j}- \mathbf t_{q,j}\right) \triangleq \chi^{\rm lb, \it r}\left(\mathbf t_{m,j}, \mathbf t_{q,j}\right). 
\end{align}
Based on this, a convex subset of non-convex constraint \eqref{P1_cons:d} can be expressed as
\begin{align}
	\chi^{\rm lb, \it r}\left(\mathbf t_{m,j}, \mathbf t_{q,j}\right) \geq D^2, \ \forall m,q\in\mathcal M, m\neq q, j\in\mathcal J.\label{P1_cons:d_sca} 
\end{align}

By replacing constraint \eqref{P1_eqv_appro_cons:b} with \eqref{P1_eqv_appro_cons:b_sca} and constraint \eqref{P1_cons:d} with \eqref{P1_cons:d_sca}, a lower bound of the optimal value
of problem \eqref{P1_eqv_appro_sub1} can be obtained by solving the following problem:
\begin{eqnarray}\label{P1_eqv_appro_sub1_sca}
	\underset{\eta, \{\mathcal T_j\}}{\max} \hspace{2mm}\eta \hspace{8mm}
	\text{s.t.} \hspace{2mm} \eqref{P1_cons:c}, \eqref{P1_eqv_appro_cons:b_sca}, \eqref{P1_cons:d_sca}. 
\end{eqnarray}
Note that problem \eqref{P1_eqv_appro_sub1_sca} is a convex quadratically constrained quadratic program (QCQP), which can be solved to optimality using standard solvers such as CVX \cite{2004_S.Boyd_cvx}. 

\subsubsection{Optimizing $\left\lbrace \mathcal R_j\right\rbrace $}  For any given $\{\mathcal T_j\}$, the subproblem of optimizing $\{\mathcal R_j\}$ can be expressed as 
\begin{eqnarray}\label{P1_eqv_appro_sub2}
	\underset{\eta, \{\mathcal R_j\}}{\max} \hspace{2mm} \eta \hspace{8mm} \text{s.t.} \hspace{2mm} \eqref{P1_eqv_appro_cons:b},\eqref{cons:modulus_slack}. 
\end{eqnarray}
To facilitate the solution design, we define the following variables: $\hat{\mathbf h}^{\mathrm H}_j(\mathbf u_j) \triangleq \left[\hat{\mathbf h}^{\mathrm H}_{j,1}(\mathbf u_j),\ldots,\hat{\mathbf h}^{\mathrm H}_{j,L}(\mathbf u_j)\right] \in \mathbb C^{1\times N}$, $\hat{\mathbf G}(\mathcal T_j) \triangleq \left[\hat{\mathbf G}_{1}(\mathcal T_j),\ldots,\hat{\mathbf G}_{L}(\mathcal T_j)\right]^T \in \mathbb C^{N\times M}$,  $\mathbf \Phi_j(\mathbf u_j) \triangleq {\rm diag}\left(\hat{\mathbf h}_j^{\mathrm H}(\mathbf u_j)\right)\hat{\mathbf G}(\mathcal T_j) \in \mathbb C^{N\times M}$, and $\mathbf v_j \triangleq \big[\mathrm{diag}(\mathbf \Theta_{1,j})^T, \ldots, 
\mathrm{diag}(\mathbf \Theta_{L,j})^T \big]^T \in \mathbb{C}^{N\times 1}$. With these definitions, the term $\left\|\sum_{\ell=1}^L\hat{\mathbf h}_{\ell,j}^{\mathrm H}(\mathbf u_j)\mathbf \Theta_{\ell,j}\hat{\mathbf G}_\ell(\mathcal T_j)\right\|^2$ in constraint \eqref{P1_eqv_appro_cons:b} can be rewritten as 
\begin{align}         
& \left\|\sum_{\ell=1}^L\hat{\mathbf h}_{\ell,j}^{\mathrm H}(\mathbf u_j)\mathbf \Theta_{\ell,j}\hat{\mathbf G}_\ell(\mathcal T_j)\right\|^2 = \left\|\hat{\mathbf h}_{k}^{\mathrm H}(\mathbf u_j)\hat{\mathbf \Theta}_j\hat{\mathbf G}(\mathcal T_j)\right\|^2 \nonumber\\
& \hspace{2.5cm} = \left\|\mathbf v_j^{\mathrm H}\mathbf \Phi_j(\mathbf u_j)\right\|^2 \triangleq \mathbf v_j^{\mathrm H}\mathbf Q_j(\mathbf u_j)\mathbf v_j,
\end{align}
where $\mathbf Q_j(\mathbf u_j)\triangleq \mathbf \Phi_j(\mathbf u_j)\mathbf \Phi_j^{\mathrm H}(\mathbf u_j) \in \mathbb C^{N\times N}$. Accordingly, problem \eqref{P1_eqv_appro_sub2} can be recast as
\begin{subequations}\label{P1_eqv_appro_sub2_eqv}
	\begin{eqnarray}
		&\hspace{-1cm} \underset{\eta, \{\mathbf v_j\}}{\max}& \hspace{-2mm}\eta \\
		&\hspace{-1cm}\text{s.t.}& \hspace{-5mm} \bar P\mathbf v_j^{\mathrm H}\mathbf Q_j(\mathbf u_j)\mathbf v_j + C_{\mathbf u_j} \geq \eta, \ \forall \mathbf u_j \in \mathcal G_j, j\in\mathcal J, \label{P1_eqv_appro_sub2_eqv_cons:b}\\
		&& \hspace{-5mm} \left|\left[\mathbf v_j\right]_n\right| \leq 1, \ \forall  j\in\mathcal J, n\in\mathcal N. \label{P1_eqv_appro_sub2_eqv_cons:c}
	\end{eqnarray}
\end{subequations}
Note that each constraint in \eqref{P1_eqv_appro_sub2_eqv_cons:b} defines a super-level set of the convex quadratic function $\mathbf v_j^{\mathrm H}\mathbf Q_j(\mathbf u_j)\mathbf v_j$, and is therefore non-convex. Nevertheless, this structure enables the use of the SCA technique. Specifically, by replacing $\mathbf v_j^{\mathrm H}\mathbf Q_j(\mathbf u_j)\mathbf v_j$ with its first-order Taylor expansion-based lower bound, denoted by $\Omega_j^{\rm lb, \it r}\left(\mathbf v_j,\mathbf u_j\right)$, constraint \eqref{P1_eqv_appro_sub2_eqv_cons:b} can be approximated as
\begin{align}\label{P1_eqv_appro_cons:b_sca_2}
	\bar P\Omega_j^{\rm lb, \it r}\left(\mathbf v_j,\mathbf u_j\right) + C_{\mathbf u_j} \geq \eta, \ \forall \mathbf u_j \in \mathcal G_j, j\in\mathcal J,  
\end{align}
where $\Omega_j^{\rm lb, \it r}\left(\mathbf v_j,\mathbf u_j\right) \triangleq 2{\rm Re}\left\lbrace \left(\mathbf v_j^r\right)^{\mathrm H}\mathbf Q_j(\mathbf u_j)\mathbf v_j\right\rbrace - \left(\mathbf v_j^r\right)^{\mathrm H}\mathbf Q_j(\mathbf u_j)\mathbf v_j^r$, with $\mathbf v_j^r$ denoting the given local point in the $r$-th iteration. Based on this approximation, problem \eqref{P1_eqv_appro_sub2_eqv} can be approximated by the following convex QCQP:
\begin{eqnarray}\label{P1_eqv_appro_sub2_eqv2_sca}
	\underset{\eta, \{\mathbf v_j\}}{\max} \hspace{2mm} \eta \hspace{8mm}
	\text{s.t.} \hspace{2mm} \eqref{P1_eqv_appro_cons:b_sca_2}, \eqref{P1_eqv_appro_sub2_eqv_cons:c}. 
\end{eqnarray} 
This problem can be efficiently solved by off-the-shelf solvers such as CVX \cite{2004_S.Boyd_cvx}, and its optimal value serves as a lower bound for that of problem  \eqref{P1_eqv_appro_sub2_eqv}, and hence also of problem \eqref{P1_eqv_appro_sub2}. 

\begin{algorithm}[!t]  
    \caption{Proposed AO algorithm for problem (P1)} \label{Alg1}  
	\begin{algorithmic}[1]
		\STATE Initialize $\left\lbrace \left\lbrace \mathbf t_{m,j}^0\right\rbrace, \left\lbrace \mathbf \Theta_{\ell,j}^0\right\rbrace\right\rbrace$ and set $r = 0$.  
		\REPEAT 
		\STATE Calculate $\left\lbrace{\rm Re}\left\lbrace b_{\ell,\ell',j}(\mathbf u_j)\right\rbrace\right\rbrace$, $\left\lbrace \nabla Z_{\ell,\ell'}(\mathbf t_{m,j}^r)\right\rbrace $, $\left\lbrace \nabla^2 Z_{\ell,\ell'}(\mathbf t_{m,j})\right\rbrace $, and $\left\lbrace \psi_{\ell,\ell',m,j}\right\rbrace$. 
		\STATE Update $\left\lbrace \mathbf t_{m,j}^{r+1}\right\rbrace$ by solving problem \eqref{P1_eqv_appro_sub1_sca} with given $\left\lbrace\left\lbrace \mathbf t_{m,j}^r\right\rbrace, \left\lbrace \mathbf \Theta_{\ell,j}^r\right\rbrace\right\rbrace$.  
		\STATE Calculate $\left\lbrace\mathbf Q_j(\mathbf u_j)\right\rbrace$ and define $\mathbf v_j^r \triangleq \big[\mathrm{diag}(\mathbf \Theta_{1,j}^r)^T, \ldots, 
		\mathrm{diag}(\mathbf \Theta_{L,j}^r)^T \big]^T$, $\forall j$. 
		\STATE Update $\left\lbrace\mathbf v_j^{r+1}\right\rbrace$ by solving problem \eqref{P1_eqv_appro_sub2_eqv2_sca} with given $\left\lbrace\mathbf t_{m,j}^{r+1}, \left\lbrace \mathbf v_j^r\right\rbrace\right\rbrace$, and set $\mathbf \Theta_{\ell,j}^{r+1} = \mathrm{Diag}\left(\mathbf v_j^{r+1} \left(\sum_{k=1}^{\ell-1}(N_k + 1) : \sum_{k=1}^\ell N_k \right)\right)$, $\forall \ell,j$. 
		\vspace{-3mm}
		\STATE $r \leftarrow r + 1$. 
		\UNTIL The fractional increase of the objective value drops below a threshold $\epsilon > 0$.
		\STATE  Set $\left[ \hat{\mathbf v}_j\right]_n = \left[\mathbf v_j^r\right]_n/\left|\left[\mathbf v_j^r\right]_n\right|$, set $\hat{\mathbf \Theta}_{\ell,j} = \mathrm{Diag}\left(\hat{\mathbf v}_j \left(\sum_{k=1}^{\ell-1} N_k + 1 : \sum_{k=1}^\ell N_k \right)\right)$, and compute $\left\lbrace \mathbf w_{\mathbf u_j}\right\rbrace$ via \eqref{equ:beam_solu} with given $\left\lbrace \left\lbrace \mathbf t_{m,j}^r\right\rbrace, \left\lbrace \hat{\mathbf \Theta}_{\ell,j}\right\rbrace\right\rbrace$. 
		\STATE Output $\left\lbrace \left\lbrace\mathbf w_{\mathbf u_j}\right\rbrace \left\lbrace \mathbf t_{m,j}^r\right\rbrace, \left\lbrace \hat{\mathbf \Theta}_{\ell,j}\right\rbrace\right\rbrace$ as a suboptimal solution of problem (P1). 
	\end{algorithmic} 
\end{algorithm}

\subsubsection{Overall Algorithm: Convergence and Complexity Analysis}
The overall procedure is summarized in Algorithm \ref{Alg1}. The convergence of is ensured by the non-decreasing objective value across iterations and the existence of an upper bound. After the convergence, since the obtained $\left\lbrace\mathbf v_j^r \right\rbrace$ may not strictly satisfy the unit-modulus constraint, we normalize each entry as $\left[ \hat{\mathbf v}_j\right]_n = \left[\mathbf v_j\right]_n/\left|\left[ \mathbf v_j\right]_n\right|$, $\forall j\in\mathcal J, n\in\mathcal N$, to enforce the constraint without affecting other feasibility conditions. 

Next, we analyze the complexity of the proposed algorithm. The dominant computational cost per iteration arises from the updates of $\left\lbrace\mathcal T_j\right\rbrace$ and $\left\lbrace\mathcal R_j\right\rbrace$. For notational convenience, let $G \triangleq \sum_{j=1}^J \left|\mathcal G_j\right|$ denote the total number of candidate locations across all the target areas. For the update of $\left\lbrace\mathcal T_j\right\rbrace$, the complexities of calculating $\left\lbrace{\rm Re}\left\lbrace b_{\ell,\ell',j}(\mathbf u_j)\right\rbrace\right\rbrace$, $\left\lbrace \nabla Z_{\ell,\ell'}(\mathbf t_{m,j}^r)\right\rbrace$, $\left\lbrace \nabla^2 Z_{\ell,\ell'}(\mathbf t_{m,j})\right\rbrace $, and $\left\lbrace \psi_{\ell,\ell',m,j}\right\rbrace$, as well as solving the QCQP in \eqref{P1_eqv_appro_sub1_sca} are $\mathcal O\left(LG\sum_{\ell=1}^LN_\ell\right)$, $\mathcal O\left(L^2MJ\right)$, $\mathcal O\left(L^2MJ\right)$, $\mathcal O\left(L^2MJ\right)$, and $\mathcal O\left(\sqrt{M^2J+G}\left(M^4J^3+M^3JG\right)\right)$ \cite{2014_K.wang_complexity}, respectively. For the update of $\left\lbrace\mathcal R_j\right\rbrace$, the complexities of calculating $\left\lbrace\mathbf Q_j(\mathbf u_j)\right\rbrace$ and solving the QCQP in \eqref{P1_eqv_appro_sub2_eqv2_sca} are $\mathcal O(N^2MG)$ and $\mathcal O\left(\sqrt{NG+NJ}\left(N^4J^2G+N^3J^3\right)\right)$ \cite{2014_K.wang_complexity}, respectively. Combining all components, the overall per-iteration complexity of the proposed algorithm is $\mathcal{O} \big( LG\sum_{\ell=1}^L N_\ell + L^2MJ + \sqrt{M^2J + G}(M^4J^3 + M^3JG) + N^2MG + \sqrt{NG + NJ}(N^4J^2G + N^3J^3)\big)$. 

\subsection{Application of the General Framework to (P2) and (P3)} 
Problems (P2) and (P3) share the same structure as (P1) and can be regarded as its special cases with different objectives and constraints. For instance, (P2) corresponds to solving (P1) under the additional convex constraint $\mathbf \Theta_1 = \mathbf \Theta_2 = \ldots = \mathcal R_j$, or equivalently, $\mathbf v_1 = \mathbf v_2 = \ldots = \mathbf v_J$ in the vectorized form. However, solving (P2) via (P1) introduces extra overhead from the added constraint. Since (P2) has fewer variables and constraints, it is more efficient to solve it directly. The solution framework for (P1) can be readily adapted to (P2) and (P3) by simplifying the variable dimensions and constraint sets while preserving the core principles (e.g., AO, surrogate function design, and SCA). For brevity, we omit the detailed procedures for (P2) and (P3). 

Furthermore, similar to the complexity analysis of the algorithm proposed for (P1) in the previous subsection, the per-iteration complexity of the algorithm for (P2) is approximately $\mathcal{O} \big( LG\sum_{\ell=1}^L N_\ell + L^2MJ + \sqrt{M^2J + G}(M^4J^3 + M^3JG) + N^2MG + N^{4.5}G^{1.5}\big)$, while that for (P3) is approximately $\mathcal{O} \big( LG\sum_{\ell=1}^L N_\ell + L^2M + \sqrt{M^2 + G}(M^4 + M^3G) + N^2MG + N^{4.5}G^{1.5}\big)$.

\begin{figure}[!t]
	\centering
	\includegraphics[width=0.8\columnwidth]{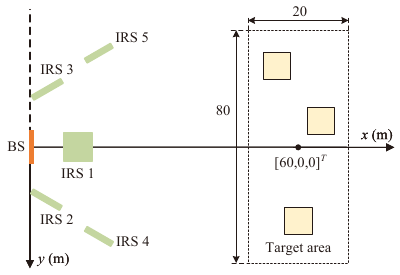}
	\caption{Simulation setup (top view). The coordinates of the reference points of the five IRSs are given by $[5,0,12]^T$, $[0,12,5]^T$, $[0,-12,5]^T$, $[10,25,5]^T$, and $[10,-25,5]^T$ in m, respectively.} \label{Fig:simulation_setup}
\end{figure} 

\vspace{-2mm}
\section{Simulation Results}\label{Sec_simulation}
In this section, we numerically evaluate the performance of the proposed schemes. Fig. \ref{Fig:simulation_setup} illustrates the top view of the considered simulation setup. Note that although five IRSs are depicted, not all are active in every simulation. The target areas are randomly generated without overlap within a rectangular region defined by $x \in [50, 70]$ m, $y \in [-40, 40]$ m, and $z = 0$ m, where each area is a square of size $5~\mathrm{m} \times 5~\mathrm{m}$ and is uniformly sampled at one-meter intervals to generate candidate locations. IRS 1 represents a horizontally oriented IRS (mounted on an uncrewed aerial vehicle (UAV) in Fig. \ref{Fig:system_model} for illustration), which more generally refers to an IRS deployed parallel to the ground at a higher position (e.g., on UAVs, indoor ceilings, rooftops, or canopies). Positioned along the central axis of the target-area region, IRS 1 establishes a more favorable BS-IRS-user geometry, leading to more balanced coverage compared with other vertically oriented IRSs (e.g., on building facades or indoor walls). The moving region for the transmit MAs is set as $\mathcal C = [-\frac{A}{2},\frac{A}{2}] \times [-\frac{A}{2},\frac{A}{2}]$ \cite{2023_Lipeng_Modeling}. The system operates at a carrier frequency of 3 GHz, corresponding to a wavelength of $\lambda = 0.1$ m \cite{2024_Zhenyu_uplink}, and the reference channel power gain at 1 m is given by $C_0 = (\lambda/4\pi)^2$. The path-loss exponents are set to $2.2$ for IRS-related links and $3.5$ for direct links, and the Rician factors of IRS-related links are set to $3$ dB \cite{2020_Xidong_NOMA}. Unless otherwise specified, we set the BS transmit power as $P = 40$ dBm, the number of transmit MAs as $M = 4$, the noise power as $\sigma^2 = -90$ dBm \cite{2020_Shuowen_Capacity}, the minimum inter-MA spacing as $D = \lambda/2$ \cite{2023_Wenyan_MIMO}, and the number of target areas as $J = 3$. Without loss of generality, we assume that all IRSs are equipped with the same number of elements, denoted by $N_{\rm e}$, i.e., $N_\ell = N_{\rm e}$, $\forall \ell \in \mathcal L$. The values of the moving region size $A$, the number of employed IRSs $L$, and the number of elements per IRS $N_{\rm e}$ will be specified individually in the simulation cases. \looseness=-1

For comparison purposes, we consider the following two benchmark schemes: 1) \textbf{FPA-(area-adaptive IRS)}: the BS employs FPA-based uniform linear arrays comprising $M$ antennas with spacing $\lambda/2$, while the IRS phase shifts are adaptively adjusted for different areas; 2) \textbf{FPA-staIRS}: the BS employs FPA-based uniform linear arrays comprising $M$ antennas with spacing $\lambda/2$, while the IRSs maintain the static phase shifts configured only once during installation. 

\begin{figure}[!t]
	\centering
	\includegraphics[scale=0.64]{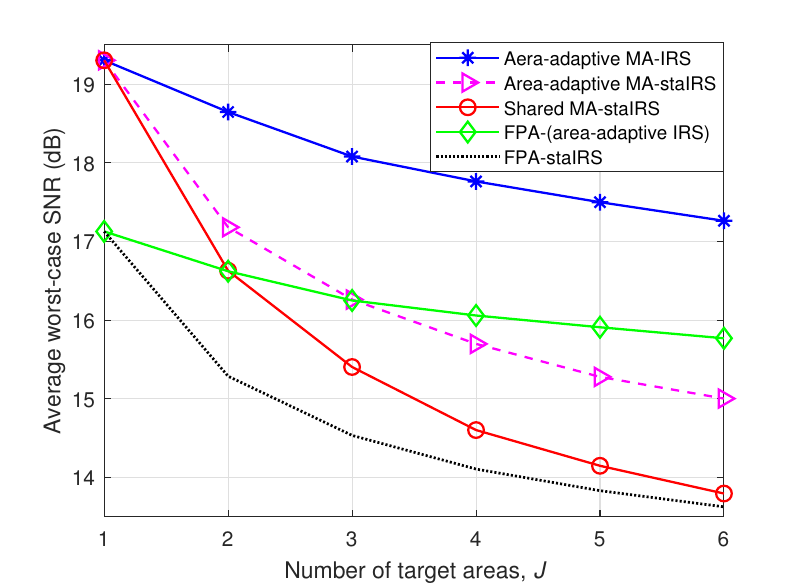}
	\caption{Worst-case SNR versus number of target areas.} \label{Fig:SNR_vs_J}
	\vspace{-2mm}
\end{figure}

In Fig.~\ref{Fig:SNR_vs_J}, we plot the worst-case SNR versus the number of target areas $J$ for $A=5\lambda$, $L=3$, and $N_{\rm e}=20$. Firstly, the worst-case SNR decreases monotonically with $J$ for all schemes, since more target areas enlarge the set of candidate locations, introduce additional constraints, and shrink the feasible region for optimization. Secondly, the area-adaptive MA-IRS scheme consistently achieves the highest SNR for all $J$, as it exploits the most spatial degrees of freedom by jointly adapting MA positions and IRS phase shifts to each area. Thirdly, when $J=1$, the three MA-based schemes perform identically, as do the two FPA-based benchmarks, because there is no distinction between adaptive and shared MA positioning or between adaptive and static IRSs in a single-area case. Fourthly, although the area-adaptive MA-staIRS scheme initially outperforms the FPA-area-adaptive IRS scheme, its advantage diminishes and is eventually reversed as $J$ increases, due to the limited flexibility of static IRSs in multi-area scenarios. As $J$ increases, static IRSs struggle to enhance the channels to all candidate locations, and with far fewer transmit MAs than IRS elements (4 versus 60), the adaptability of MAs alone cannot sustain performance superiority. In contrast, area-adaptive IRSs paired with FPAs can better tailor signal enhancement to each area, yielding superior performance for large $J$. Finally, as $J$ increases, the performance gain of the shared MA-staIRS scheme over the FPA-staIRS scheme declines, since shared MAs face increasing difficulty in balancing signal enhancement across more target areas.

\begin{figure}[!t]
	\centering
	\includegraphics[scale=0.64]{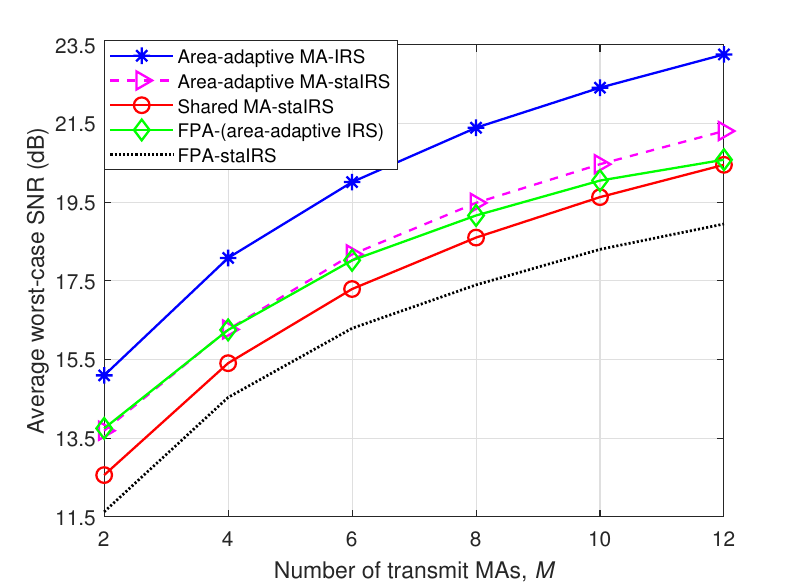}
	\caption{Worst-case SNR versus number of transmit MAs.} \label{Fig:SNR_vs_M}
	\vspace{-3mm}
\end{figure}

\begin{figure}[!t]
	\centering
	\includegraphics[scale=0.64]{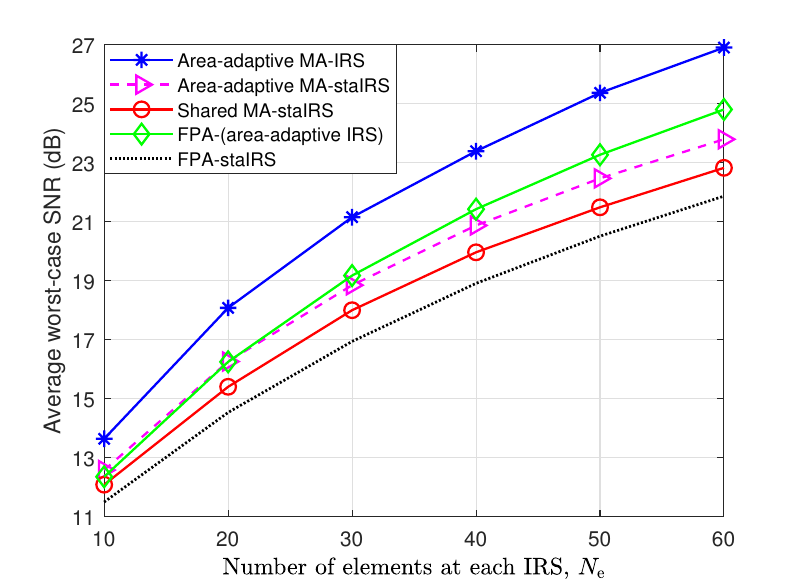}
	\caption{Worst-case SNR versus number of elements at each IRS.} \label{Fig:SNR_vs_elements}
	\vspace{-2mm}
\end{figure}

\begin{figure}[!ht]
	\centering
	\subfigure[$N_{\rm e} = 20$.]{\label{Fig:SNR_vs_L}
		\includegraphics[scale=0.64]{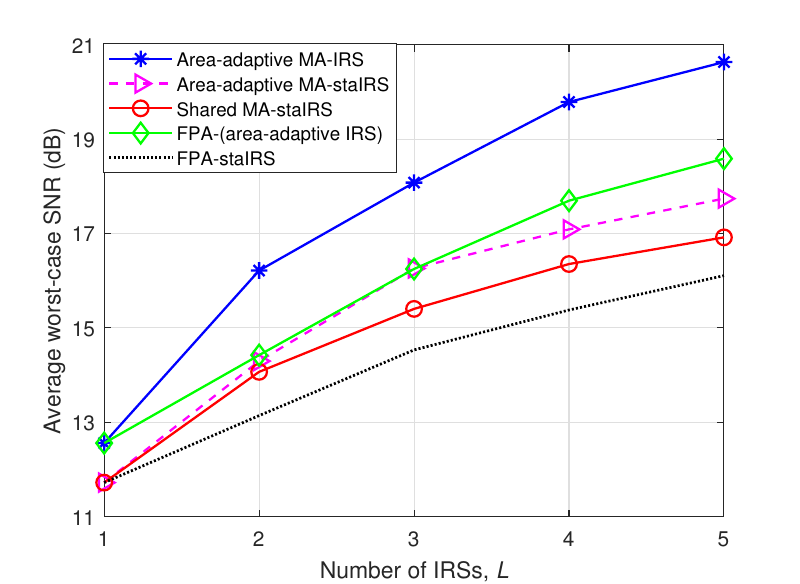}}
		
	\vspace{-2mm}
	\subfigure[$N = 120$ and $N_{\rm e} = N/L$.]{\label{Fig:SNR_vs_L_fixN}
		\includegraphics[scale=0.64]{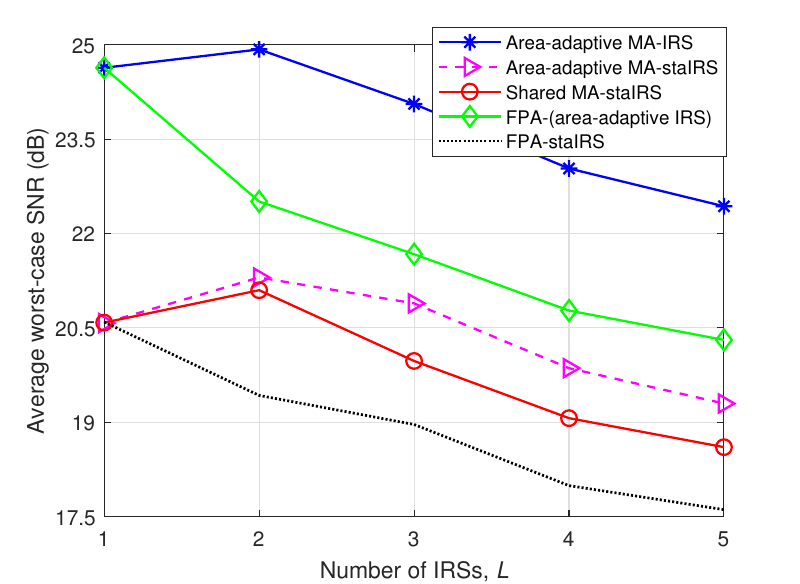}}
	\caption{Worst-case SNR versus number of IRSs.}
	\label{Fig:SNR_vs_L_two}
	\vspace{-2mm}
\end{figure}

Fig. \ref{Fig:SNR_vs_M} depicts the worst-case SNR performance versus the number of transmit MAs, under the setting of $A = 8\lambda$, $L = 3$, and $N_{\rm e} = 20$. We observe that the area-adaptive MA-staIRS scheme initially underperforms the FPA-(area-adaptive IRS) scheme at small $M$ values (e.g., $M=2$ versus $60$ IRS elements), as its limited spatial processing capability is insufficient to rival the benefits of dynamically tuned IRS elements. Nevertheless, as $M$ increases, the area-adaptive MA-staIRS scheme improves steadily and eventually achieves superior performance, benefiting from the enhanced spatial degrees of freedom provided by additional MAs. Meanwhile, the shared MA-staIRS scheme, though constrained by common MA positioning, also narrows its performance gap with the FPA-(area-adaptive IRS) scheme as $M$ grows, indicating that more MAs can partially offset the lack of area-specific adaptability.

Fig.~\ref{Fig:SNR_vs_M} shows the worst-case SNR versus the number of transmit MAs for $A=8\lambda$, $L=3$, and $N_{\rm e}=20$. The area-adaptive MA-staIRS scheme initially underperforms the FPA-area-adaptive IRS scheme at small $M$ (e.g., $M=2$ versus 60 IRS elements) due to its limited spatial processing capability. As $M$ increases, its performance improves steadily and eventually surpasses that of the FPA-area-adaptive IRS scheme, benefiting from the additional spatial degrees of freedom. The shared MA-staIRS scheme, though constrained by common MA positioning, also narrows its performance gap with the FPA-area-adaptive IRS scheme as $M$ grows, indicating that more MAs can partially offset the lack of area-specific adaptability. These results show that MA-based systems with static IRSs require a sufficient number of transmit antennas to overcome their initial disadvantage relative to FPA-based systems with area-adaptive IRSs, underscoring the need for proper MA-IRS dimensioning in practice. 

In Fig. \ref{Fig:SNR_vs_elements}, we examine the impact of the number of elements per IRS on the worst-case SNR performance with $A = 5\lambda$ and $L = 3$. As expected, the worst-case SNR increases with $N_{\rm e}$ across all schemes, as more IRS elements provide stronger passive beamforming gain. While both static and area-adaptive IRS configurations benefit from this increase, the gain is more pronounced in schemes with adaptive IRSs. This explains not only the observed crossover between the area-adaptive MA-staIRS and FPA-(area-adaptive IRS) schemes, but also the widening performance gap between the area-adaptive MA-IRS and area-adaptive MA-staIRS schemes.

\begin{figure*}[!ht]
	\centering
	\vspace{-2mm}
	\subfigure[$C_{\rm tot} = 120c_{\rm e}$.]{\label{Fig:scheme1_vs_M_rhochange}
		\includegraphics[scale=0.64]{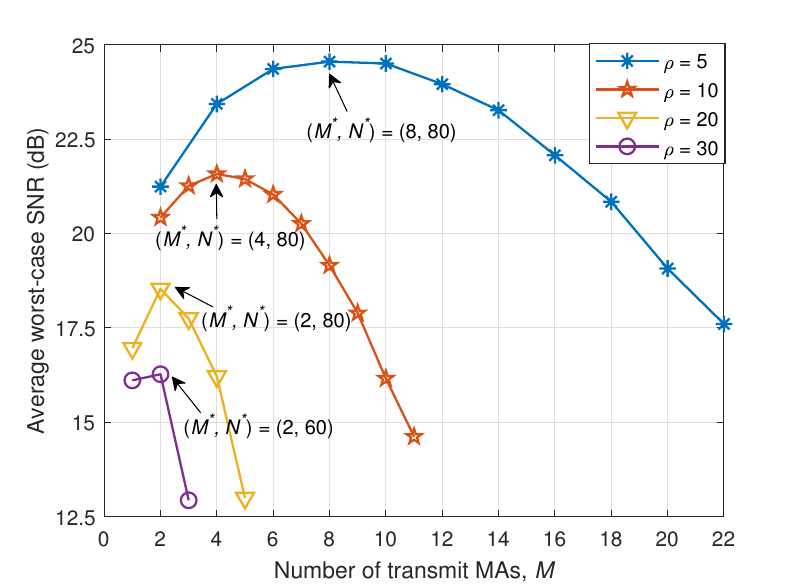}}
	\subfigure[$\rho = 20$.]{\label{Fig:scheme1_vs_M_costchange_rho20}
		\includegraphics[scale=0.64]{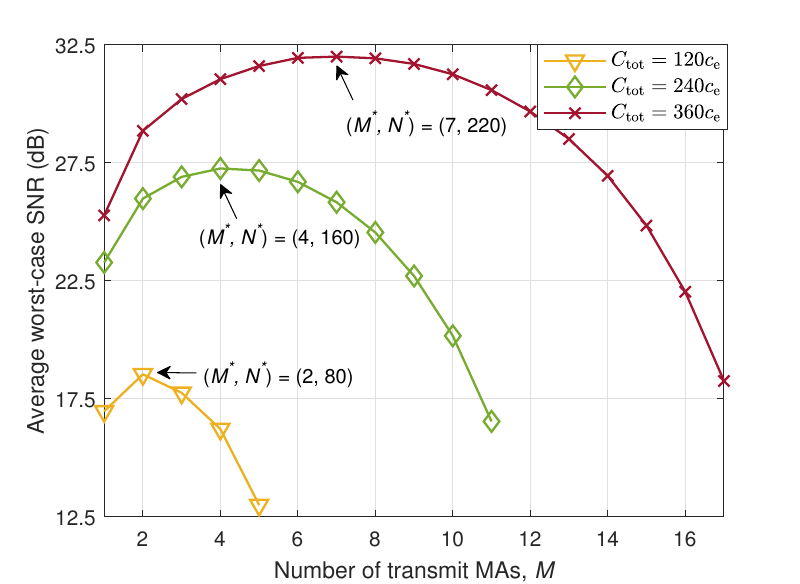}}
	\caption{Worst-case SNR of the area-adaptive MA-IRS scheme versus number of transmit MAs.}
	\label{Fig:scheme1_vs_M}
	\vspace{-2mm}
\end{figure*}

In Fig. \ref{Fig:SNR_vs_L_two}, we plot the worst-case SNR versus the number of IRSs for $A = 5\lambda$ under two setups: (a) fixed number of elements per IRS, $N_{\rm e} = 20$; (b) fixed total elements, $N = 120$, evenly distributed across IRSs so that $N_{\rm e} = N/L$. From both Figs. \ref{Fig:SNR_vs_L} and \ref{Fig:SNR_vs_L_fixN}, it is observed that when $L=1$, the area-adaptive MA-IRS and FPA-(area-adaptive IRS) schemes yield the same worst-case SNR, as do the area-adaptive MA-staIRS, shared MA-staIRS, and FPA-staIRS schemes. This shows that with a single IRS, performance is mainly determined by whether IRS beamforming is adaptive or static, while the form of transmitter architecture (MAs or FPAs) has a negligible impact. This is consistent with results in \cite{2024_Weidong_MAIRS}, where theory and simulation show that the MA gain over FPAs disappears under far-field LoS BS–IRS channels. However, when $L\geq 2$, the superiority of MAs over FPAs becomes apparent, as more IRSs introduce richer multipath and greater spatial channel variation, enabling MAs to be positioned in high-gain regions for higher received SNR. 

Moreover, in Fig. \ref{Fig:SNR_vs_L}, the worst-case SNR of all schemes is observed to increase with $L$, as more IRSs provide extra signal paths (spatial diversity gain) and more reflecting elements (passive beamforming gain). These two effects also jointly explain the non-monotonic gap between the FPA-(area-adaptive IRS) and area-adaptive MA-staIRS schemes: the gap first narrows due to the increasing spatial diversity of the area-adaptive MA-staIRS design, and then widens as the FPA-(area-adaptive IRS) scheme benefits more from the growing passive beamforming gain, attributed to its area-adaptive IRS configuration. 
In contrast, Fig. \ref{Fig:SNR_vs_L_fixN} shows that the worst-case SNR of the two FPA-based benchmark schemes decreases with $L$, because a fixed total number of elements must be shared among IRSs with different effectiveness. Specifically, IRS 1 with the shortest BS distance and a central position relative to the target areas, offers the best coverage.  Redistributing its elements to less effective IRSs (IRS 2–5) reduces passive beamforming gain, degrading performance. For the three MA-based schemes, however, the worst-case SNR first increases from $L = 1$ to $L = 2$ as spatial diversity gain from MA position adaptation outweighs element redistribution loss, but declines for $L \geq 3$ when further partitioning to ineffective IRSs makes passive beamforming loss dominant. 

In Fig. \ref{Fig:scheme1_vs_M}, we evaluate the worst-case SNR performance of the area-adaptive MA-IRS scheme under a fixed total cost of transmit MAs and IRS elements, with $A = 8\lambda$, $L = 2$, and $N_{\rm e} = N/L$. Let $c_{\rm e}$ denote the unit cost of an IRS element and $c_{\rm M} = \rho c_{\rm e}$ the cost of a single transmit MA, where $\rho$ is the cost ratio between an MA and an IRS element. The total cost is thus $C_{\rm tot} = c_{\rm M}M + c_{\rm e}N = \rho c_{\rm e}M + c_{\rm e}N$. This relative cost modeling approach follows the principle used in \cite{2021_Jiangbin_cost}.
Fig. \ref{Fig:scheme1_vs_M_rhochange} examines the effect of $\rho$ with $C_{\rm tot} = 120c_{\rm e}$, while Fig.  \ref{Fig:scheme1_vs_M_costchange_rho20} studies $C_{\rm tot}$ variation with $\rho = 20$. Both show a non-monotonic trend: the worst-case SNR first rises with $M$, peaks at an optimal $M^*$, then declines, revealing an optimal $(M^*, N^*)$ balance between active and passive resources. Although increasing $M$ boosts spatial diversity and beamforming flexibility, over-allocating to MAs (at the cost of IRS elements) severely impairs passive beamforming. Additionally, $M^*$ exhibits two key trends: 1) under fixed $C_{\rm tot}$, it decreases with $\rho$, reflecting the need to prioritize IRS elements when MAs become relatively expensive; 2) under fixed $\rho$, it increases approximately linearly with $C_{\rm tot}$, as higher budgets permit more MAs without severely compromising IRS deployment. 

An empirical resource allocation rule is observed as: $\frac{M^*}{N^*} \approx \frac{a}{\rho}$, where $a$ captures active-passive trade-off. Simulation results show that $a$ typically stabilizes around $0.5$ for moderate $\rho$ and budgets (e.g., $\rho = 5, 10, 20$ with $C_{\rm tot} = 120c_{\rm e}$; $\rho = 20$ with $C_{\rm tot} = 240c_{\rm e}$). Moreover, $a$ slightly rises with abundant budgets (e.g., $a\approx0.636$ for $\rho=20$, $C_{\rm tot}=360c_{\rm e}$), and can even reach $1.0$ in high-$\rho$ scenarios with limited budgets (e.g., $a = 1.0$ for $\rho=30$, $C_{\rm tot}=120c_e$). This indicates that, even when MAs are relatively expensive, the existence of direct links preserves the marginal benefit of active beamforming, leading the system to allocate a non-negligible fraction of resources to MAs. Importantly, the relation $\frac{M^*}{N^*}\approx\frac{a}{\rho}$ remains independent of $C_{\rm tot}$ and is mainly governed by $\rho$, implying that the system inherently maintains a $\rho$-dependent balance between active and passive resources across different budget levels. Besides, by combining $\frac{M^*}{N^*} \approx \frac{a}{\rho}$ with $C_{\rm tot} = \rho c_{\rm e}M + c_{\rm e}N$, we derive: $M^* \approx \frac{a}{a+1}\frac{C_{\rm tot}}{\rho c_{\rm e}}$. This provides a practical guideline for optimal MA count under a given total cost. 

\section{Conclusion}\label{Sec_conclusion}
In this paper, we proposed three practical coverage-enhancement schemes for IRS-aided MA systems: the area-adaptive MA-IRS scheme, the area-adaptive MA-staIRS scheme, and the shared MA-staIRS scheme. To maximize the worst-case SNR across all target areas, we formulated corresponding optimization problems that jointly optimized the MA positioning, IRS configuration, and BS beamforming for each proposed scheme. Despite the non-convexity and implicit objective functions of the three optimization problems, we proposed a general algorithmic framework to efficiently obtain suboptimal solutions for each. Numerical results confirmed the superiority of the proposed MA-based schemes over FPA-based counterparts across different IRS configurations, with the area-adaptive MA-IRS scheme consistently delivering the highest worst-case SNR performance. Moreover, in scenarios with static IRS configurations, deploying a modest number of transmit MAs is critical for the system to surpass FPA-based systems with area-adaptive IRSs, underscoring the importance of balanced MA-IRS deployment strategies. Finally, under a fixed total cost constraint, practical guidelines for resource allocation were identified, revealing that the optimal MA-IRS quantity ratio is mainly determined by their relative cost. This provides a useful reference for cost-efficient dimensioning of MAs and IRSs in practical deployments. 

While this paper has focused on coverage enhancement in designated target areas, in practice, the BS typically also needs to serve users located outside these areas. In such cases, an IRS configuration optimized for the target areas may unintentionally influence the service quality in other service areas. A valuable extension is therefore to incorporate these other areas into the optimization design, for example, through time-domain switching, unified optimization over enlarged area sets, or threshold-based protection. We regard this as a promising direction for future research.

\appendix[Proof of Proposition \ref{prop1}]
By substituting $\mathbf w_{\mathbf u_j}^* = \frac{\left( \sum_{\ell=1}^L\mathbf h_{\ell,j}^{\mathrm H}(\mathbf u_j)\mathbf \Theta_{\ell,j}\mathbf G_\ell(\mathcal T_j) + \mathbf f_{\mathbf u_j}\right)^{\mathrm H}}{\left\|\sum_{\ell=1}^L\mathbf h_{\ell,j}^{\mathrm H}(\mathbf u_j)\mathbf \Theta_{\ell,j}\mathbf G_\ell(\mathcal T_j) + \mathbf f_{\mathbf u_j}\right\|}$ into $\gamma_j^{(1)}\left(\mathcal T_j, \mathcal R_j, \mathbf w_{\mathbf u_j}^*,\mathbf u_j\right)$, we obtain the closed-form expression shown in \eqref{equ:gamma_1_expansion} at the top of the next page, where $\hat{\mathbf h}^{\mathrm H}_{\ell,j}(\mathbf u_j)$ and $\hat{\mathbf G}_{\ell}(\mathcal T_j)$ are defined as in Proposition \ref{prop1}. 
\begin{figure*}[!t]
	\begin{align}\label{equ:gamma_1_expansion}
		& \gamma_j^{(1)}\left(\mathcal T_j, \mathcal R_j, \mathbf w_{\mathbf u_j}^*,\mathbf u_j\right) = \bar P\left\| \sum_{\ell=1}^L\mathbf h_{\ell,j}^{\mathrm H}(\mathbf u_j)\mathbf \Theta_{\ell,j}\mathbf G_\ell(\mathcal T_j) + \mathbf f_{\mathbf u_j}\right\|^2 \nonumber\\
		& \hspace{1.5cm} = \bar P\left\|\sum_{\ell=1}^L\left(\hat{\mathbf h}^{\mathrm H}_{\ell,j}(\mathbf u_j)+\sqrt{\frac{C_0d_{\ell,j}^{-\alpha_{\ell,j}}(\mathbf u_j)}{\kappa_{\ell,j} + 1}}\tilde{\mathbf h}_{\ell,j}^{\mathrm H}(\mathbf u_j)\right)\mathbf \Theta_{\ell,j}\left(\hat {\mathbf G}_\ell(\mathcal T_j) + \sqrt{\frac{C_0d_{\rm \ell}^{-\alpha_\ell}}{\kappa_\ell + 1}}\tilde{\mathbf G}_\ell(\mathcal T_j)\right) + \mathbf f_{\mathbf u_j}\right\|^2\nonumber\\ 
		& \hspace{1.5cm} = \bar P\left\| \sum_{\ell=1}^{L} \left[ \underbrace{\hat{\mathbf{h}}_{\ell,j}^{\mathrm H}(\mathbf u_j) \mathbf{\Theta}_{\ell,j}\hat{\mathbf{G}}_\ell (\mathbf{t}_j)}_{\mathbf x_{\ell,1}} + \underbrace{\sqrt{\frac{C_0 d_\ell^{-\alpha_\ell}}{\kappa_\ell + 1}} \hat{\mathbf{h}}_{\ell,j}^{\mathrm H}(\mathbf u_j) \mathbf{\Theta}_{\ell,j} \tilde{\mathbf{G}}_\ell(\mathcal T_j)}_{\mathbf x_{\ell,2}} + \underbrace{\sqrt{\frac{C_0 d_{\ell,j}^{-\alpha_{\ell,j}} (\mathbf{u}_j)}{\kappa_{\ell,j} + 1}} \tilde{\mathbf{h}}_{\ell,j}^{\mathrm H}(\mathbf u_j) \mathbf{\Theta}_{\ell,j} \hat{\mathbf{G}}_\ell (\mathcal T_j)}_{\mathbf x_{\ell,3}} \right.\right.  \nonumber\\
		& \hspace{2cm} \left. \left. + \underbrace{C_0 \sqrt{\frac{d_{\ell,j}^{-\alpha_{\ell,j}}(\mathbf{u}_j) d_\ell^{-\alpha_\ell} }{(\kappa_{\ell,j} + 1)(\kappa_\ell + 1)}} \tilde{\mathbf{h}}_{\ell,j}^{\mathrm H}(\mathbf u_j) \mathbf{\Theta}_{\ell,j} \tilde{\mathbf{G}}_\ell(\mathcal T_j)}_{\mathbf x_{\ell,4}}\right] + \mathbf f_{\mathbf u_j}\right\|^2. 
	\end{align}
	\hrulefill
\end{figure*}
Then, we focus on the subsequent derivation of $\mathbb E_{\mathcal B}\left[\left\| \left(\sum_{\ell=1}^L\sum_{k=1}^4\mathbf x_{\ell,k} + \mathbf x_{\ell,2} + \mathbf x_{\ell,3} + \mathbf x_{\ell,4}\right) + \mathbf f_{\mathbf u_j} \right\|^2\right]$. Owing to the zero-mean and mutual independence of $\tilde{\mathbf G}_\ell(\mathcal T_j)$, $\tilde{\mathbf h}_{\ell,j}^{\mathrm H}(\mathbf u_j)$, and $\tilde{\mathbf f}_{\mathbf u_j}$, all cross terms involving them in the expansion of the squared norm have zero expectation. We thus obtain $\mathbb E_{\mathcal B}\left[\left\|\left(\sum_{\ell=1}^L\mathbf x_{\ell,1} + \mathbf x_{\ell,2} + \mathbf x_{\ell,3} + \mathbf x_{\ell,4}\right) + \mathbf f_{\mathbf u_j}\right\|^2\right] = \left\|\sum_{\ell=1}^L\mathbf x_{\ell,1}\right\|^2  + \sum_{\ell=1}^L\Big(\mathbb E_{\mathcal B}\left[\left\|\mathbf x_{\ell,2}\right\|^2\right] + \mathbb E_{\mathcal B}\left[\left\|\mathbf x_{\ell,3}\right\|^2\right] + \mathbb E_{\mathcal B}\left[\left\|\mathbf x_{\ell,4}\right\|^2 \right] \Big) + \mathbb E_{\mathcal B}\left[\left\|\mathbf f_{\mathbf u_j}\right\|^2\right]$,
where 
\begin{align}\label{equ:gamma_1_expansion_terms}
	\mathbb E_{\mathcal B}\left[\left\|\mathbf x_{\ell,2}\right\|^2\right] & \! = \!\frac{C_0 d_\ell^{-\alpha_\ell}}{\kappa_\ell + 1} \!\times\! C_0 d_{\ell,j}^{-\alpha_{\ell,j}}(\mathbf u_j) \frac{\kappa_{\ell,j}}{\kappa_{\ell,j} + 1} M \left\|  \overline{\mathbf{h}}_{\ell,j}\right\|^2 \nonumber\\
	& = \Lambda_{\mathbf u_j}\kappa_{\ell,j}MN_\ell, \nonumber\\
	\mathbb E_{\mathcal B}\left[\left\|\mathbf x_{\ell,3}\right\|^2\right] & = \frac{C_0 d_{\ell,j}^{-\alpha_{\ell,j}} (\mathbf{u}_j)}{\kappa_{\ell,j} + 1}\!\times \! C_0 d_\ell^{-\alpha_\ell} \frac{\kappa_\ell}{\kappa_\ell + 1} \left\| \overline{\mathbf{G}}_\ell \right\|_F^2\nonumber\\
	& = \Lambda_{\mathbf u_j}\kappa_{\ell}MN_\ell, \nonumber\\
	\mathbb E_{\mathcal B}\left[\left\|\mathbf x_{\ell,4}\right\|^2\right] & = \frac{C_0^2d_{\ell,j}^{-\alpha_{\ell,j}}(\mathbf u_j)d_{\rm \ell}^{-\alpha_\ell}}{\left(\kappa_{\ell,j}+1\right)\left(\kappa_\ell+1\right)} \times MN_\ell = \Lambda_{\mathbf u_j}MN_\ell, \nonumber\\
	\mathbb E_{\mathcal B}\left[\left\|\mathbf f_{\mathbf u_j}\right\|^2\right] & = C_0d_{\mathbf u_j}^{-\alpha_j}\mathbb E\left[\left\|\tilde{\mathbf f}_{\mathbf u_j}\right\|^2\right] = C_0d_{\mathbf u_j}^{-\alpha_j}M,
\end{align}
where $\Lambda_{\mathbf u_j} \triangleq \frac{C_0^2d_{\ell,j}^{-\alpha_{\ell,j}}(\mathbf u_j)d_{\rm \ell}^{-\alpha_\ell}}{\left(\kappa_{\ell,j}+1\right)\left(\kappa_\ell+1\right)}$, as defined in Proposition \ref{prop1}.
Combining the above, we arrive at 
\begin{align}
    &\hspace{-1.2mm}\mathbb E_{\mathcal B}\!\left[\gamma_j^{(1)}\!\left(\mathcal T_j, \mathcal R_j, \mathbf w_{\mathbf u_j}^*,\mathbf u_j\right)\right] \!=\! \bar P\left\|\sum_{\ell=1}^L\hat{\mathbf h}_{\ell,j}^{\mathrm H}(\mathbf u_j)\mathbf \Theta_{\ell,j}\hat{\mathbf G}_\ell(\mathcal T_j)\right\|^2 \nonumber\\
	& + \bar P\left( \sum_{\ell=1}^L\left[\left(\kappa_{\ell,j} + \kappa_\ell + 1\right)\Lambda_{\mathbf u_j}MN_{\ell}\right] + C_0d_{\mathbf u_j}^{-\alpha_j}M\right).  
\end{align}
This completes the proof of Proposition \ref{prop1}.
\vspace{-2mm}

\bibliographystyle{IEEEtran}
\bibliography{ref}
\flushend

\end{document}